\documentclass[a4paper,11pt]{article}

\def\al{\alpha}
\def\be{\beta}
\def\ga{\gamma}
\def\de{\delta}
\def\ep{\epsilon}
\def\ve{\varepsilon}

\def\et{\eta}
\def\th{\theta}

\def\ka{\kappa}
\def\la{\lambda}

\def\rh{\rho}

\def\si{\sigma}

\def\ta{\tau}

\def\ph{\phi}

\def\ch{\chi}
\def\ps{\psi}
\def\om{\omega}
\def\Ga{\Gamma}
\def\De{\Delta}

\def\La{\Lambda}

\def\Om{\Omega}

\def\cE{{\cal E}}
\def\cl{{\cal L}}
\def\cL{{\cal L}}

\def\mn{{\mu\nu}}

\def\fr#1#2{{{#1} \over {#2}}}
\def\half{{\textstyle{1\over 2}}}
\def\quar{{\textstyle{1\over 4}}}
\def\frac#1#2{{\textstyle{{#1}\over {#2}}}}

\def\vev#1{\langle {#1}\rangle}

\def\lsim{\mathrel{\rlap{\lower4pt\hbox{\hskip1pt$\sim$}}
    \raise1pt\hbox{$<$}}}
\def\gsim{\mathrel{\rlap{\lower4pt\hbox{\hskip1pt$\sim$}}
    \raise1pt\hbox{$>$}}}
\def\sqr#1#2{{\vcenter{\vbox{\hrule height.#2pt
         \hbox{\vrule width.#2pt height#1pt \kern#1pt
         \vrule width.#2pt}
         \hrule height.#2pt}}}}

\def\prt{\partial}

\def\etal{{\it et al.}}

\def\pt#1{\phantom{#1}}

\def\ol#1{\overline{#1}}

\def\nsc#1#2#3{\om_{#1}^{{\pt{#1}}#2#3}}
\def\lsc#1#2#3{\om_{#1#2#3}}

\def\lulsc#1#2#3{\om_{#1\pt{#2}#3}^{{\pt{#1}}#2}}

\def\vb#1#2{e_{#1}^{{\pt{#1}}#2}}
\def\ivb#1#2{e^{#1}_{{\pt{#1}}#2}}
\def\uvb#1#2{e^{#1#2}}
\def\lvb#1#2{e_{#1#2}}

\def\ss{$s^{\mu\nu}$}
\def\tt{$t^{\ka\la\mu\nu}$}
\def\uu{$u$}

\def\sss{s^{\mu\nu}}
\def\ttt{t^{\ka\la\mu\nu}}

\newcommand{\beq}{\begin{equation}}
\newcommand{\eeq}{\end{equation}}
\newcommand{\bea}{\begin{eqnarray}}
\newcommand{\eea}{\end{eqnarray}}
\newcommand{\bit}{\begin{itemize}}
\newcommand{\eit}{\end{itemize}}
\newcommand{\rf}[1]{(\ref{#1})}

\def\lrDmu{\stackrel{\leftrightarrow}{D_\mu}}
\def\lrDnu{\stackrel{\leftrightarrow}{D^\nu}}

\begin{document}

\title{Overview of the SME: Implications and Phenomenology of Lorentz Violation}

\author{Robert Bluhm \\ 
Colby College, Waterville, ME 04901, USA \\ 
}

\maketitle             

\begin{abstract}
The Standard Model Extension (SME) provides the most general observer-independent
field theoretical framework for investigations of Lorentz violation.
The SME lagrangian by definition contains all Lorentz-violating interaction terms
that can be written as observer scalars and that 
involve particle fields in the Standard Model and gravitational fields in a
generalized theory of gravity.
This includes all possible terms that could arise from a process of
spontaneous Lorentz violation in the context of a more fundamental theory,
as well as terms that explicitly break Lorentz symmetry.
An overview of the SME is presented, 
including its motivations and construction.
Some of the theoretical issues arising in the case of 
spontaneous Lorentz violation are discussed,
including the question of what happens to the Nambu-Goldstone modes
when Lorentz symmetry is spontaneously violated and whether a
Higgs mechanism can occur.
A minimal version of the SME in flat Minkowski spacetime
that maintains gauge invariance and power-counting renormalizability
is used to search for leading-order signals of Lorentz violation.
Recent Lorentz tests in QED systems are examined, 
including experiments with photons, particle and atomic experiments,
proposed experiments in space,
and experiments with a spin-polarized torsion pendulum.
\end{abstract}

\section{Introduction}
\label{intro}

It has been 100 years since Einstein published his first papers on
special relativity
\cite{ae}.
This theory is based on the principle of Lorentz invariance,
that the laws of physics and the speed of light are the same in
all inertial frames.
A few years after Einstein's initial work,
Minkowski showed that a new spacetime geometry emerges from special relativity.
In this context,
Lorentz symmetry is an exact spacetime symmetry
that maintains the form of the Minkowski metric 
in different Cartesian-coordinate frames.

In the years 1907-1915,
Einstein developed the general theory of relativity as a new theory of gravity.
In general relativity,
spacetime is described in terms of a metric that
is a solution of Einstein's equations.
The geometry is Riemannian,
and the physics is invariant
under general coordinate transformations.
Lorentz symmetry, on the other hand,
becomes a local symmetry.
At each point on the spacetime manifold,
local coordinate frames can be found
in which the metric becomes the Minkowski metric.
However, the choice of the local frame is not unique,
and local Lorentz transformations provide the link between
physically equivalent local frames.

The Standard Model (SM) of particle physics is a fully relativistic theory.
The SM in Minkowski spacetime is invariant under global Lorentz transformations,
whereas in a Riemannian spacetime the particle interactions must remain
invariant under both general coordinate transformations and local Lorentz
transformations.
Particle fields are also invariant under gauge transformations.
Exact symmetry under local gauge transformations leads to the existence
of massless gauge fields, such as the photon.
However, spontaneous breaking of local gauge symmetry in the
electroweak theory involves the Higgs mechanism,
in which the gauge fields can acquire a mass.

Classical gravitational interactions can be described in a form
analogous to gauge theory by using a vierbein formalism
\cite{uk}.
This also permits a straightforward treatment of fermions
in curved spacetimes.
Covariant derivatives of tensors in the local Lorentz frame
involve introducing the spin connection.
In a Riemann spacetime with zero torsion,
the spin connection is not an independendent field,
but rather is a prescribed function of the vierbein
and its derivatives.
However, a natural generalization is to treat
the spin connection components as independent
degrees of freedom.
The resulting geometry is a Riemann-Cartan spacetime,
which has nonvanishing torsion
\cite{rcst}.
In a Riemann-Cartan spacetime, 
the associated field strengths for the vierbein and
spin connection are the curvature and torsion tensors.
The usual Riemann spacetime of general relativity is
recovered in the zero-torsion limit.
Similarly,
if the curvature tensor vanishes,
the spacetime reduces to Minkowski spacetime.

The combination of the SM and Einstein's classical gravitational
theory provides a highly successful description of nature.
However,
since Einstein's theory is not a quantum theory,
it is expected that it will ultimately be superseded by a more
fundamental theory that will hold at the quantum level.
Candidate quantum gravity theories include string theory and
loop quantum gravity.
The appropriate scale where gravity and quantum physics
are expected to meet up is the Planck scale,
$m_P \simeq 10^{19}$ GeV.

Finding experimental confirmation of a quantum theory of gravity by
doing experiments at the Planck scale, however, is not practical.
Instead,
an alternative approach can be adopted in which one looks 
for small Planck-suppressed effects of new physics
that might be observable in high-precision experiments.
For this idea to hold,
any new effect would have to be one that cannot be mimicked by
known conventional processes in the SM or conventional gravity theory.
One possible signal fulfilling this requirement is to look
for Planck-suppressed signatures of Lorentz violation in high-precision experiments.

Detection of such a violation of relativity theory would clearly be a dramatic
indication of new physics, 
presumably coming from the Planck scale.
This idea is not merely speculative because it has been shown that
mechanisms in both string theory 
\cite{ks,kps}
and quantum gravity 
\cite{qg}
can lead to violations of Lorentz symmetry.
However,
these theories are not yet sufficiently developed in such a way
that allows testable predictions to be made at a definite 
(quantifiable) scale at low energies.

Nonetheless,
progress can still be made using effective field theory.
To be realistic,
an effective field theory would have to contain both the SM
and general relativity together with any higher-order couplings
between them.
It should also maintain coordinate (or observer) independence.
In full generality,
the gravity sector could include additional fields such as torsion 
that are not a part of Einstein's general relativity.
This would permit more general geometries,
including a Riemann-Cartan spacetime.

The general effective field theory of this type incorporating
arbitrary observer-independent Lorentz violation is called
the Standard-Model Extension (SME)
\cite{kpo,ck,akgrav}.
The SME lagrangian by definition contains all observer-scalar terms 
consisting of products of SM and general gravitational fields with
each other as well as with additional couplings that introduce
violations of Lorentz symmetry.
In principle,
there are an infinity of terms in the SME,
including nonrenormalizable terms of arbitrarily high dimension.

To investigate low-energy experiments,
where the leading-order signals of Lorentz violation
are of primary interest,
it is often advantageous to work with a subset of the full SME,
which includes only a finite number of terms.
One subset in particular,
referred to as the minimal SME,
restricts the theory to power-counting renormalizable and
gauge-invariant terms.
In recent years,
the Lorentz-violating coefficients in the minimal SME have been 
adopted by experimentalists as the standard for reporting bounds
on Lorentz violation.
Since many of the low-energy experiments involve only
electromagnetic interactions between charged particles and photons,
it often suffices to define a minimal QED sector of the SME.

This paper is intended as an overview in the context of the SME
of some recent theoretical and phenomenological investigations of Lorentz violation.
The motivations for the development of the SME are presented first.
An outline of how the theory is constructed is then given.
This is followed by a discussion of some theoretical issues
that come up when Lorentz violation is due to a process
of spontaneous symmetry breaking.
In particular,
the fate of the Nambu-Goldstone modes is examined along
with the question of whether a Higgs mechanism can occur
\cite{rbak05}.
For simplicity,
this discussion is carried out in the context of a vector model known as
a bumblebee model
\cite{ks2,akgrav}.
The role of the geometry (Minkowski, Riemann, or Riemann-Cartan)
is examined as well.
To investigate phenomenology,
the minimal SME is constructed and used to examine a wide range of experiments
assuming a flat Minkowski background.
In this paper,
the focus is on high-precision tests in QED systems.
A number of recent experiments in atomic and particle 
systems are examined,
and the status of their attainable sensitivities to Lorentz violation
is reviewed.

The SME is the result of a large on-going collaboration by a group of
theorists and experimentalists most of whom have in common that they have 
at some point collaborated with Alan Kostelecky at Indiana University.
An exhaustive review covering all of this collective work,
which spans topics in field theory, gravity, astrophysics, cosmology,
as well as particle, nuclear, and atomic physics, is not possible here.
Instead, this review focuses mostly on selective recent 
topics that are of interest to the author.
It is also not possible here to give a complete list of references
on all of the work looking at possible violations or tests of relativity.
For that, other recent reviews and proceedings collections should 
be consulted as well.
See, for example, 
Refs.\ \cite{cpt123,leh01,revs}.

\section{Motivations}
\label{motivations}

Historically,
interest in Lorentz violation increased dramatically after it was discovered by
Kostelecky and Samuel in the late 1980s that mechanisms can occur in string field theory
that could cause spontaneous breaking of Lorentz symmetry
\cite{ks}.
It is this idea that ultimately led to the development of the SME,
which in turn has stimulated a variety of experimental searches for
relativity violations.

Spontaneous Lorentz violation can occur 
when a string field theory has a nonperturbative
vacuum that can lead to tensor-valued fields acquiring nonzero
vacuum expectation values (vevs), $\vev{T} \ne 0$.
As a result of this,
the low-energy effective theory contains an unlimited number of terms
of the form 
\beq
 \cl \sim \fr \la {m_P^k} \, \vev{T} \, \Ga \, \bar \ps (i \partial )^k \chi
\quad ,
\label{Lterms}
\eeq
where $k$ is an integer power, $\la$ is a coupling constant,
and $\ps$ and $\ch$ are fermion fields.
In this expression, the tensor vev $\vev{T}$ carries spacetime indices,
which are suppressed in this notation.
This vev is effectively a set of functions or constants that are fixed in a given
observer frame.
What this means is that interactions with these coefficients
can have preferred directions in spacetime or velocity (boost) dependence.
The vev coefficients therefore induce Lorentz violation.

Note that the higher-dimensional ($k>0$) derivative couplings
are expected to be balanced by additional inverse factors of a large mass scale,
which is assumed to be the Planck mass $m_P$. 
In a more complete low-energy effective theory describing fermions $\ps$ and $\ch$
there could also be other terms with additional couplings,
including possible Yukawa couplings.
A more general interaction term of the form in Eq.\ \rf{Lterms}
at order $k$ could then be written as
\beq
 \cl \sim t^{(k)} \, \Ga \, \bar \ps (i \partial )^k \chi
\quad ,
\label{Lterms2}
\eeq
where the coefficient $t^{(k)}$,
which carries spacetime indices,
absorbs all of the couplings, inverse mass factors, and the vev.
This effective coefficient acts essentially as a fixed 
background field that induces Lorentz violation.
In addition to interactions with fermions,
additional terms involving gauge-field couplings and gravitational
interactions are possible as well.
A generalization would be to include all possible contractions of
known SM and gravitational fields 
with fixed background coefficients $t^{(k)}$.

This generalization to include all arbitrary-dimension interaction terms inducing
Lorentz violation in effective field theory is the idea behind the SME
\cite{kpo,ck,akgrav}.
Note as well that each term is assumed to be an observer scalar,
with all spacetime indices contracted.
The full SME is then defined as the effective field theory obtained
when all such scalar terms are formed using SM and gravitational fields
contracted with coefficients that induce Lorentz violation.
The SME coefficients (the generalized $t^{(k)}$ factors)
are assumed to be heavily suppressed,
presumably by inverse powers of the Planck mass.
The extent of the suppression increases with order $k$.
Without a completely viable string field theory,
it is not possible to assign definite numerical values to 
these coefficients,
and clearly (as in the SM itself) there are hierarchy issues. 
However,
since no Lorentz violation has been observed in nature,
it must be that the SME coefficients are small.
Alternatively, 
one can adopt a phenomenological approach and treat the coefficients as
quantities to be bounded in experiments.
Such bounds will also constitute a measure of the sensitivity
to Lorentz violation attained in the experiment.

Interestingly,
although the SME was originally motivated from ideas in string field theory,
including the idea of spontaneous Lorentz symmetry breaking,
its relevance and usefulness extend well beyond these ideas.
In fact,
there is nothing in the SME that requires that the Lorentz-violation
coefficients emerge due to a process of spontaneous Lorentz violation.
The SME coefficients can also be viewed as due to explicit Lorentz violation
or as arising from some unknown mechanism.
Indeed,
once the philosophy of the SME is appreciated --
that it is the most general observer-independent field theory 
incorporating Lorentz violation --
then no matter what scalar lagrangian is written down 
involving known low-energy fields,
the result will be contained in the full SME.

An illustration of this comes from studying noncommutative field theory.
These are theories that have noncommuting coordinates 
\beq 
[x^\mu , x^\nu ] = i \th^{\mu\nu}
\quad .
\label{xnc}
\eeq
It has been shown that this type of geometry can occur naturally
in string theory
\cite{ncft},
and that it leads to Lorentz violation 
\cite{ncqed}.
Here, however, the mechanism leading to Lorentz violation
is in general different from that of spontaneous symmetry breaking.
Nonetheless,
the form of the effective interactions that arise
are fully contained in the SME.
The fixed parameters $\th^{\mu\nu}$,
which break the Lorentz symmetry,
act effectively so as to produce SME coefficients.
For example,
the effective field theory involving a $U(1)$ gauge field
in a noncommutative geometry includes lagrangian terms of the form
\beq
\cl \sim \fr 1 4 \, i q \, \th^{\al\be} \, F_{\al \be} \, \bar \ps
\, \ga^\mu \, D_\mu \, \ps 
\quad ,
\label{Lncterms}
\eeq
where $F_{\al \be}$ is the field strength.
Here,
as in Eq.\ \rf{Lterms}
the interaction takes the form of a scalar-valued product of
known particle fields, derivative operators, and a set
of fixed background functions.

There are a number of other examples of effective field theories
with Lorentz violation that have been put forward in recent years,
with a wide variety of motivations or ideas for symmetry breaking.
Nonetheless,
as long as the resulting theories are described by scalar lagrangians,
then they are compatible with the approach of the SME.
For example,
a model with spatial rotational invariance was used in
Ref.\ \cite{cg99} 
to study high-energy cosmic rays above the GZK cutoff.
Another example with a
higher-dimensional lagrangian giving rise to Lorentz-violating
dispersion relations was considered in 
Ref.\ \cite{mp03}.
An example involving gravitational fields
includes a parameterized set of kinetic terms for a vector field
in a theory with spontaneous Lorentz breaking
\cite{jm04}.
In all of these cases,
the lagrangian terms can be found as a subset of the full SME.

Over the years,
a number of phenomenological frameworks that involve specific types of
Lorentz violation have been developed and used extensively by experimentalists.
A sampling includes the $TH\ep\mu$ model
\cite{them},
the Robertson-Mansouri-Sexl framework
\cite{rms}, 
the PPN formalism
\cite{will},
as well as models based on kinematical breaking of Lorentz symmetry
(see Refs.\ \cite{cpt123,revs} for reviews).
In some cases,
these theories describe parameterized equations of motion 
or dispersion relations and
do not originate from a scalar lagrangian.
However,
to the extent that these models can be described by effective
field theory defined by a scalar lagrangian,
they are compatible with the SME and direct links between
their parameterizations and the SME coefficients can be obtained.

It should be noted as well,
that in addition to breaking Lorentz symmetry,
the SME also leads to violation of the discrete symmetry CPT
\cite{ks,kps}.
This symmetry is the product of charge conjugation (C),
parity (P), and time reversal (T).
According to the CPT theorem
\cite{cptthm},
a relativistic field theory describing point particles should
exactly obey CPT symmetry.
Conversely, a second theorem states that if CPT is violated
in field theory, then Lorentz symmetry must also be broken
\cite{owg}.
It then follows that any observer-independent effective field theory describing
CPT violation must also be contained within the SME.
Since CPT can be tested to very high precision in experiments
with matter and antimatter,
this opens up a whole new avenue for exploring the phenomenology
of Lorentz violation.

In summary,
the full SME is defined as the most general observer-independent theory of
Lorentz and CPT violation that contains the SM and gravity.
It thus provides a unifying framework that can be used to investigate
possible signals of Lorentz and CPT violation.
Because it contains an infinity of terms,
with an unlimited set of coefficients with spacetime indices,
it is not possible to list all of them.
However, the terms can be classified in a general way,
and a uniform notation can be developed.
It is also possible to develop subset theories of the full SME,
which contain a finite number of terms.
One subset in particular has been investigated extensively in recent experiments.
It is the minimal SME,
which is comprised of the gauge-invariant subset of terms in the full SME
with dimension four or less.

Finally,
one other remark about the SME coefficients is worth mentioning.
It is often commented these coefficients,
such as for example a nonzero vacuum vev of a tensor field generated from
a process of spontaneous Lorentz violation,
are reminiscent of the old pre-relativistic ether.
However,
the ether was thought to be a medium (with a rest frame) for light,
whereas an SME coefficient need not be thought of in this way.
The SME coefficients act effectively as background vacuum fields.
Their interactions typically select out a particular particle species.
In fact,
if that particle is not the photon, then the SME coefficient
will have no direct influence on the speed of light.
Moreover,
the SME coefficients carry tensor indices and therefore have 
definite spacetime directions in any observer frame.
In the end,
while there are some similarities to the old ether,
the physical effects of the SME coefficients are significantly different.

\section{Constructing the SME}
\label{sme}

One of the defining features of the SME is that
the theory is observer independent
\cite{ck}.
It is therefore important to make clear the distinction between what
are called {\it observer} and {\it particle} Lorentz transformations.
An observer Lorentz transformation is a change of observer frame.
It can be viewed as a rotation or boost of the basis vectors
in the local frame.
The philosophy of the SME is that even with Lorentz violation,
physics must remain observer independent.
The results of an experiment should not depend on the
chosen perspective of any observer.
In contrast,
a particle Lorentz transformation is a rotation or boost
performed on an individual particle field while leaving the coordinate frame fixed.
In this case,
if there is Lorentz violation,
the physics can change.

In terms of what this means for an experiment,
the observer invariance of the SME says that
the results of a measurement cannot depend on the choice of
coordinate frame or observational perspective made by the experimenter.
On the other hand,
if Lorentz symmetry is broken,
the results of the experiment can change if the apparatus itself
is rotated or boosted in some direction,
both of which are examples of particle Lorentz transformations.

Note that this feature of the SME breaks the relativity principle,
which is a central assumption of (unbroken) relativity theory.
This principle is often stated as the equivalence of
passive and active Lorentz transformations when one 
is performed as the inverse of the other.
In the formulation of the SME, 
however,
the terms passive and active are deliberately avoided since
for one thing their usage is sometimes confused in the literature.
More importantly, though,
it is observer independence that is the physically defining feature
of the theory,
and the terminology should reflect this.
In addition,
observers need not be inactive.
The idea in the SME is that even if an observer actively changes its perspective or 
relative motion with respect to the apparatus in an experiment,
the results of measurements should remain unchanged.

A similar distinction between observer and particle transformations
can be made for general coordinate transformations performed in 
the spacetime manifold of a Riemann or Riemann-Cartan geometry
\cite{akgrav,rbak05}. 
An observer transformation is simply a change of spacetime coordinates,
which leaves the physics unchanged.
On the other hand,
a particle transformation is essentially a diffeomorphism,
which maps one point on the spacetime to another.
The change in a tensor under pullback to the original point is
given by the Lie derivative.

The full SME is defined using a vierbein formalism.
This permits a natural distinction between the spacetime manifold
and local Lorentz frames.
The vierbein $\vb \mu a$ provides
a link between the components of a tensor field
$T_{\la\mu\nu\cdots}$ on the spacetime manifold
(denoted using Greek indices)
and the corresponding components 
$T_{abc\cdots}$ in a local Lorentz frame
(denoted using latin indices).
The correspondence is given by 
\beq
T_{\la\mu\nu\cdots} 
= \vb \la a \vb \mu b \vb \nu c \cdots T_{a b c \cdots} .
\label{vier}
\eeq
In this notation,
the components of the spacetime metric are $g_{\mu\nu}$,
while in a local Lorentz frame, 
the metric takes the Minkowski form $\et_{ab}$.
A necessary condition for the vierbein is therefore
that $g_{\mu\nu} = \vb \mu a \vb \nu b \et_{ab}$.
Covariant derivatives acting on tensor fields with local
indices introduce the spin connection $\nsc \mu a b$.
For example,
\beq
D_\mu \vb \nu a = 
\prt_\mu \vb \nu a 
-\Ga^\al_{\pt\al\mu\nu} \vb \al a
+ \lulsc \mu a b \vb \nu b .
\eeq
In a Riemann spacetime where $D_\la g_{\mu\nu} = 0$,
the spin connection is not an independent field,
but rather is a prescribed function of the vierbein
and its derivatives.
However, in a Riemann-Cartan spacetime the spin
connection represents independent degrees of freedom
associated with there being nonzero torsion.

The observer independence of the SME requires that all of the terms in
the lagrangian be observer scalars under both 
general coordinate transformations and local Lorentz transformations.
This means that every spacetime index 
and every local Lorentz index  
must be fully contracted in the lagrangian.

However,
the SME is not invariant under particle diffeomorphisms and
local Lorentz transformations.
Explicitly,
a diffeomorphism maps one point on the spacetime to another.
It can be characterized infinitesimally in a coordinate basis by
the transformation
\beq
x^\mu \rightarrow x^\mu + \xi^\mu .
\label{diffeos}
\eeq
The four infinitesimal parameters $\xi^\mu$ 
comprise the diffeomorphism degrees of freedom.
On the other hand,
under an infinitesimal particle Lorentz transformation 
the field components transform
through contraction with a matrix of the form
\beq
\La^a_{\pt{a}b} \approx \de^a_{\pt{a}b} + \ep^a_{\pt{a}b} ,
\label{LLT}
\eeq
where $\ep_{ab} = - \ep_{ba}$ are the infinitesimal parameters 
carrying the six Lorentz degrees of freedom
and generating the local Lorentz group.
Evidently,
there are a total of ten relevant spacetime symmetries.

Violation of these symmetries occurs when an interaction term contains
coefficients that remain fixed under a particle transformation,
such as when as a particle rotation or boost is performed
in a background with a fixed vev.

\subsection{Minimal SME}
\label{msme}

The full SME consists of an unlimited number of observer scalar terms
consisting of contractions of SM fields, gravitational fields,
and SME coefficients.
To begin to explore phenomenology,
it makes sense to advance incrementally.
Since gauge symmetry and renormalizability are
foundations of our current understanding in particle physics,
a first increment would be to construct a subset theory
that maintains these features.
It is referred to as the minimal SME.
It will first be defined in Minkowski spacetime
and then generalized to include gravitational fields
in a Riemann-Cartan geometry.

The minimal SME,
constructed from dimension four or fewer operators,
describes the leading-order effects of Lorentz violation.
This is because the higher-dimensional terms are expected
to be suppressed by additional inverse powers of the Planck mass
compared to those in the minimal SME.
Effects involving couplings to gravitational fields are also
expected to be smaller than those involving interactions in the SM,
particularly electrodynamic interactions.
For this reason, the Lorentz tests described later on
are investigated using primarily a QED subset of the minimal 
SME in flat Minkowski spacetime.
Nonetheless, 
it should be kept in mind
that a particular type of Lorentz violation might only occur at
subleading order.
For this reason, it is important ultimately to investigate more
general tests in the context of the full SME,
including gravitational effects as well as
interactions involving higher-dimensional terms.
However,
that goes beyond the scope of this overview.

To construct the minimal SME in 
flat Minkowski spacetime
\cite{ck},
the first ingredient that must be put in is the minimal SM itself.
This consists of quark and lepton sectors,
gauge fields, and a Higgs sector.
Denote the left- and right-handed 
lepton and quark multiplets by
\beq
L_A = \left( \begin{array}{c} \nu_A \\ l_A
\end{array} \right)_L
\quad , \quad
R_A = (l_A)_R
\quad ,
\label{mults}
\eeq
\beq
Q_A = \left( \begin{array}{c} u_A \\ 
d_A \end{array} \right)_L ~~ , ~~ 
U_A = (u_A)_R ~~ , ~~ D_A = (d_A)_R
\quad , 
\eeq
where $A = 1,2,3$ labels the flavor,
with $l_A = (e, \mu, \ta)$, 
$\nu_A = (\nu_e, \nu_\mu, \nu_\ta)$,
$u_A = (u,c,t)$, and $d_A = (d,s,b)$.
The Higgs doublet is denoted by $\ph$.
The SU(3), SU(2), and U(1) gauge fields
are $G_\mu$, $W_\mu$, and $B_\mu$, respectively,
with corresponding field strengths:
$G_{\mu\nu}$, $W_{\mu\nu}$, and $B_{\mu\nu}$.
The gauge couplings are $g_3$, $g$, and $g^\prime$,
while $q$ denotes the electric charge.
The Yukawa couplings are $G_L$, $G_U$, $G_D$.

The relevant sectors in the SM lagrangian are:
\beq
\cl_{\rm lepton} = 
\half i \overline{L}_A \ga^{\mu} \lrDmu L_A
+ \half i \overline{R}_A \ga^{\mu} \lrDmu R_A
\label{smlepton}
\quad ,
\eeq
\beq
\cl_{\rm quark} =
\half i \overline{Q}_A \ga^{\mu} \lrDmu Q_A
+ \half i \overline{U}_A \ga^{\mu} \lrDmu U_A
+ \half i \overline{D}_A \ga^{\mu} \lrDmu D_A
\quad ,
\label{smquark}
\eeq
\beq
\hskip -0.1cm
\cl_{\rm Yukawa} = 
- \left[ (G_L)_{AB} \overline{L}_A \ph R_B
+ (G_U)_{AB} \overline{Q}_A \ph^c U_B 
+ (G_D)_{AB} \overline{Q}_A \ph D_B
\right],
\label{smyukawa}
\eeq
\beq
\cl_{\rm Higgs} =
(D_\mu\ph)^\dagger D^\mu\ph 
+\mu^2 \ph^\dagger\ph - \fr \la {3!} (\ph^\dagger\ph)^2
\quad ,
\label{smhiggs}
\eeq
\beq
\cl_{\rm gauge} =
-\half {\rm Tr} (G_{\mu\nu}G^{\mu\nu})
-\half {\rm Tr} (W_{\mu\nu}W^{\mu\nu})
-\frac 1 4 B_{\mu\nu}B^{\mu\nu} ,
\label{smgauge}
\eeq
where $D_\mu$ are gauge-covariant derivatives.

The SME introduces additional lagrangian terms that
are contractions of these SM fields with the SME coefficients .
The SME coefficients are constrained by 
the requirement that the lagrangian be hermitian.
For an SME coefficient with an even number of spacetime indices,
the pure trace component is irrelevant because it maintains Lorentz invariance.
Such coefficients may therefore be taken as traceless.

In the fermion sector of the minimal SME,
four sets of terms can be classified 
according to whether they involve leptons or quarks
and whether CPT is even or odd.
They are 
\beq
\cl^{\rm CPT-even}_{\rm lepton} = 
\half i (c_L)_{\mu\nu AB} \overline{L}_A \ga^{\mu} \lrDnu L_B
+ \half i (c_R)_{\mu\nu AB} \overline{R}_A \ga^{\mu} \lrDnu R_B
\label{lorvioll} 
\, ,
\eeq
\beq
\cl^{\rm CPT-odd}_{\rm lepton} = 
- (a_L)_{\mu AB} \overline{L}_A \ga^{\mu} L_B
- (a_R)_{\mu AB} \overline{R}_A \ga^{\mu} R_B
\label{cptvioll}
\quad ,
\eeq
\bea
\cl^{\rm CPT-even}_{\rm quark} &=& 
\half i (c_Q)_{\mu\nu AB} \overline{Q}_A \ga^{\mu} \lrDnu Q_B
+ \half i (c_U)_{\mu\nu AB} \overline{U}_A \ga^{\mu} \lrDnu U_B 
\nonumber\\ &&
+ \half i (c_D)_{\mu\nu AB} \overline{D}_A \ga^{\mu} \lrDnu D_B
\quad ,
\label{lorviolq}
\eea
\bea
\cl^{\rm CPT-odd}_{\rm quark} &=& 
- (a_Q)_{\mu AB} \overline{Q}_A \ga^{\mu} Q_B
- (a_U)_{\mu AB} \overline{U}_A \ga^{\mu} U_B 
\nonumber\\ &&
- (a_D)_{\mu AB} \overline{D}_A \ga^{\mu} D_B
\quad .
\label{cptviolq}
\eea
In these expressions,
the coefficients $a_\mu$ have dimensions of mass,
while $c_{\mu\nu}$ are dimensionless and traceless.

The couplings between fermions and the Higgs field
are all CPT even and are
\bea
\cl^{\rm CPT-even}_{\rm Yukawa} &=& 
- \half 
\left[
(H_L)_{\mu\nu AB} \overline{L}_A \ph \si^{\mu\nu} R_B
+(H_U)_{\mu\nu AB} \overline{Q}_A \ph^c \si^{\mu\nu} U_B 
\right.
\nonumber\\ &&
\quad\quad\quad\quad \left.
+(H_D)_{\mu\nu AB} \overline{Q}_A \ph \si^{\mu\nu} D_B
\right]
\quad ,
\label{loryukawa}
\eea
where the SME coefficients $H_{\mu\nu}$ are 
dimensionless and antisymmetric.

The Higgs sector itself can be CPT even or odd.
The terms are 
\bea
\cl^{\rm CPT-even}_{\rm Higgs} &=&
\half (k_{\ph\ph})^{\mu\nu} (D_\mu\ph)^\dagger D_\nu\ph 
-\half (k_{\ph B})^{\mu\nu} \ph^\dagger \ph B_{\mu\nu}
\nonumber\\ &&
\quad\quad\quad\quad
-\half (k_{\ph W})^{\mu\nu} \ph^\dagger W_{\mu\nu} \ph 
\quad ,
\label{lorhiggs}
\eea
\beq
\cl^{\rm CPT-odd}_{\rm Higgs}
= i (k_\ph)^{\mu} \ph^{\dagger} D_{\mu} \ph  
\quad .
\label{cpthiggs}
\eeq
The dimensionless coefficient $k_{\ph\ph}$ 
can have symmetric real
and antisymmetric imaginary parts.
The other coefficients 
have dimensions of mass.

The gauge sector consists of
\bea
\cl^{\rm CPT-even}_{\rm gauge} &=&
-\half (k_G)_{\ka\la\mu\nu} {\rm Tr} (G^{\ka\la}G^{\mu\nu})
-\half (k_W)_{\ka\la\mu\nu} {\rm Tr} (W^{\ka\la}W^{\mu\nu})
\nonumber\\ &&
\quad\quad\quad
-\frac 1 4 (k_B)_{\ka\la\mu\nu} B^{\ka\la}B^{\mu\nu}
\quad ,
\label{lorgauge}
\eea
\bea
\cl^{\rm CPT-odd}_{\rm gauge} &=&
(k_3)_\ka \ep^{\ka\la\mu\nu} 
{\rm Tr} (G_\la G_{\mu\nu} + \frac 2 3 i g_3 G_\la G_\mu G_\nu)
\nonumber\\ &&
+ (k_2)_\ka \ep^{\ka\la\mu\nu} 
{\rm Tr} (W_\la W_{\mu\nu} + \frac 2 3 i g W_\la W_\mu W_\nu)
\nonumber\\ &&
+ (k_1)_\ka \ep^{\ka\la\mu\nu} B_\la B_{\mu\nu} 
+ (k_0)_\ka B^\ka 
\quad .
\label{cptgauge}
\eea
The coefficients $k_{G,W,B}$ are dimensionless, 
have the symmetries of the Riemann tensor,
and have a vanishing double trace.
The coefficients $k_{1,2,3}$ are real and have dimensions of mass,
while $k_0$ is also real and has dimensions of mass cubed.
Note that if any of these CPT-odd terms appear in the theory,
they would generate instabilities associated with negative 
contributions to the energy.
For this reason,
the coefficients $k_{0,1,2,3}$ are assumed to vanish.
Interestingly,
it appears that no radiative corrections in the SME
appear to generate nonzero values for these coefficients,
at least to one loop.

It is also important to realize that some of the SME terms
can be eliminated by field redefinitions
\cite{ck,kleh,cm02}.
For example,
some of the terms involving the coefficients $a_{L,R,Q,U,D}$
can be eliminated by position-dependent field-phase redefinitions.
Another example is that certain terms involving the coefficients 
$c_{L,R,Q,U,D}$ can be absorbed by the
terms involving the coefficients $H_{L,U,D}$
through field-normalization redefinitions.
In particular,
what this means is that while a field theory can be written down
that ostensibly has explicit Lorentz violation,
it is sometimes the case that there are no physical effects
because the theory is equivalent through field redefinitions
to a Lorentz-invariant theory.

Clearly,
there are a number of additional theoretical issues that are relevant to the
construction of the SME as a consistent low-energy field theory incorporating
Lorentz violation.
These include a more in-depth discussion of 
the nature of field theory with Lorentz violation
(including quantization of the theory)
\cite{ck}, 
issues related to causality
\cite{kleh}, 
the possibility of additional extensions including for
example supersymmetry
\cite{bk02}, 
renormalization
\cite{klp}, 
electroweak symmetry breaking
\cite{ck}, 
radiative corrections
\cite{rcs}, 
spacetime variations of couplings
\cite{klpe},
etc.
It is not possible to describe all of these issues here.
The interested reader is referred to the original papers.

\subsection{Gravity Sector}

The gravity sector of the SME has been discussed in
Ref.\ \cite{akgrav},
and the minimal theory (dimension four or fewer terms) 
has been explicitly constructed.
A vierbein formalism is used,
which gives the theory a close parallel to gauge theory.
Lorentz breaking occurs due to the presence of SME coefficients,
which remain fixed
under particle Lorentz transformations in a local frame.
In this case,
the SME coefficients carry Latin indices,
e.g., $b_a$ for a vector,
with respect to the local basis set.
The conversion to spacetime coordinates is implemented
by the vierbein,
giving, e.g., $b_\mu = \vb \mu a b_a$.
The lagrangian can then be written in terms of fields
and SME coefficients defined on the spacetime manifold.
A natural (though not required) assumption is that
the SME coefficients are smooth functions over the manifold.
It is not necessary to require that they be covariantly constant.
In fact, defining covariantly constant tensors over a manifold
places stringent topological constraints on the geometry.
One simplifying assumption,
which could occur naturally in the context of
spontaneous Lorentz breaking,
is to assume that the SME coefficients are constants in the local frame.
However,
again,
this is not a requirement in the formulation of the SME theory.

To construct the minimal SME including gravity,
the first step is to incorporate gravitational fields into the usual SM.
This is done by rewriting all of the terms in 
Eqs.\ \rf{smlepton} through \rf{smgauge} with fields and gamma matrices
defined with respect to the local frame (using Latin indices).
The vierbein is then used to convert these terms over to the spacetime manifold.
Factors of the determinant of the vierbein $e$ are included as well
so that integration of the lagrangian density (giving the action)
is covariant.
Derivatives are understood as well to be both spacetime
and gauge covariant.
With these changes,
Eq.\ \rf{smlepton}, for example, becomes
\bea
\cl_{\rm lepton} = 
\half i e \ivb \mu a \ol{L}_A \ga^{a} \lrDmu L_A
+ \half i e \ivb \mu a \ol{R}_A \ga^{a} \lrDmu R_A .
\label{smlepton2}
\eea
The other terms for the quark, Yukawa, Higgs, and gauge
sectors follow a similar pattern.

The Lorentz-violating SME terms constructed from SM fields 
are obtained in a similar way.
The various particle sectors can again be divided between
CPT odd and even contributions.
Each of the terms in Eqs.\ \rf{lorvioll} to \rf{cptgauge}
is then written using local indices and vierbeins,
which convert the equations over to the spacetime manifold.
As an example,
Eq.\ \rf{lorvioll} becomes 
\bea
\cl^{\rm CPT-even}_{\rm lepton} &=& 
-\half i (c_L)_{\mu\nu AB} e \ivb \mu a \ol{L}_A \ga^{a} \lrDnu L_B
\nonumber\\ &&
\quad\quad - \half i (c_R)_{\mu\nu AB} e \ivb \mu a \ol{R}_A \ga^{a} \lrDnu R_B .
\label{lorvioll2}
\eea
The remaining equations follow the same pattern.

The pure-gravity sector of the minimal SME consists of a
Lorentz-invariant gravity sector and a Lorentz-violating sector.
The Lorentz-invariant lagrangian consists of terms that are products of
the gravitational fields.
In the general case,
this includes terms constructed from curvature, torsion,
and covariant derivatives.
Einstein's gravity (with or without a cosmological term)
would be a special case in this sector.

The Lorentz-violating lagrangian terms in the 
gravity sector of the minimal SME 
are constructed by combining the SME coefficients 
with gravitational field operators
to produce an observer scalar under
local Lorentz transformations
and general coordinate transformations.
These consist of products of the vierbein, the spin connection,
and their derivatives, 
but for simplicity 
they can be written in terms of
the curvature, torsion, and covariant derivatives.
The minimal case (up to dimension four) has the form:
\bea
{\cl}_{e, \om}^{\rm LV}
&=&
e (k_T)^{\la\mu\nu} T_{\la\mu\nu}
+ e (k_R)^{\ka\la\mu\nu} R_{\ka\la\mu\nu}
+ e (k_{TT})^{\al\be\ga\la\mu\nu} T_{\al\be\ga} T_{\la\mu\nu}
\nonumber\\ &&
\quad\quad\quad\quad
+ e (k_{DT})^{\ka\la\mu\nu} D_{\ka} T_{\la\mu\nu} .
\label{lvlag}
\eea
The SME coefficients in this expression have 
the symmetries of the associated
Lorentz-violating operators.
All except $(k_T)^{\la\mu\nu}$,
which has dimensions of mass,
are dimensionless.

The Lorentz-violating sector introduces additional gravitational
couplings that can have phenomenological consequences,
including effects on cosmology, black holes, gravitational radiation,
and post-Newtonian physics.
As a starting point for a phenomenological investigation of
the gravitational consequences of Lorentz violation,
it is useful to write down the Riemannian limit of
the minimal SME gravity sector.
It is given as
\cite{akgrav}
\bea
S_{e, \om, \La}
&=& 
 \fr 1 {2\ka}\int d^4 x 
[ e(1-u)R -2e\La
\nonumber\\
&&
\pt{\fr 1 {2\ka} \int d^4 x}
+ e \sss R_{\mu\nu}
+ e \ttt R_{\ka\la\mu\nu}].
\label{Ract}
\eea
The SME coefficient $(k_R)^{\ka\la\mu\nu}$ has been expanded into
coefficients \ss, \tt, \uu\ 
that distinguish the effects 
involving the Riemann, Ricci, and scalar curvatures.
The coefficients \ss\ have the symmetries of the Ricci tensor,
while \tt\ has those of the Riemann tensor.
Taking tracelessness conditions into account,
there are 19 independent components.

Another useful limit is the QED subset of the SME.
This extension in Minkowski space has been used extensively to
investigate high-precision experimental tests of Lorentz symmetry
in atomic and particle systems.
Generalizing to include gravity involves introducing
additional vierbein-fermion couplings as well as a pure-gravity sector.
These additional terms can then be investigated for potential
signals of Lorentz violation due to gravitational effects
in high-precision experiments.

A full treatment of the gravity sector of the SME should include
looking at the energy momentum tensor, Einstein's equations,
and consistency relations between these stemming from, for example,
the Bianchi identities.
These types of issues are described in depth in
Ref.\ \cite{akgrav}.
Interestingly,
a difference between theories with explicit versus spontaneous
breaking of Lorentz symmetry is found.
In a generic Riemann-Cartan theory with explicit breaking of 
Lorentz symmetry,
the Bianchi identities are not consistent with the
the covariant conservation laws and equations of motion.
On the other hand,
if Lorentz symmetry is spontaneously broken,
the problem is evaded.

\section{Spontaneous Lorentz Violation}
\label{spv}

One of the original motivations for developing the SME
was that mechanisms in string theory suggest that local
Lorentz symmetry might be spontaneously broken
\cite{ks}.
While the full SME describes any observer-independent Lorentz
violation at the level of effective field theory,
one important special case is when Lorentz symmetry is
spontaneously broken.
This provides an elegant mechanism 
in which the symmetry holds dynamically, 
but is broken (or hidden) by the solutions of the theory.
The lagrangian and equations of motion still respect the symmetry,
however; the vacuum values of the fields do not.
Tensor-valued fields acquire nonzero vevs which
have definite spacetime directions,
thereby breaking the symmetry under boosts and rotations.

There are certain theoretical issues that arise when the
Lorentz violation is due to spontaneous symmetry breaking.
This section examines some of these issues,
in particular,
what the fate is of the Nambu-Goldstone modes when
Lorentz symmetry is spontaneously broken.

In gauge theory,
it is well known that 
when a continuous global symmetry is spontaneously broken,
massless Nambu-Goldstone (NG) modes appear 
\cite{ng}.
If instead the broken symmetry is local,
then a Higgs mechanism can occur in which
the gauge bosons become massive  
\cite{hm}.
The question naturally arises as to what the fate of the NG modes
is when Lorentz symmetry is spontaneously broken and whether a Higgs mechanism can
occur for the case of local Lorentz symmetry (as in a theory with gravity).

This question has recently been addressed in detail in 
Ref.\ \cite{rbak05}.
A generic analysis of theories with spontaneous Lorentz breaking
was carried out in Riemann-Cartan spacetime and in the limiting cases of
Riemann and Minkowski spacetime.
A number of general features were found.

First, a connection between spontaneous breaking of local Lorentz symmetry
and diffeomorphisms was found to hold.
This occurs because when the vierbein takes a vacuum value,
which for simplicity we can take as its value in a Minkowski background,
$\vb \mu a = \de_\mu^a$,
then if a local tensor acquires a fixed vev, e.g., $b_a$ for the case of a vector,
which breaks local Lorentz symmetry,
then the associated spacetime vector $b_\mu$ as given by contraction with the
vierbein also acquires a fixed vev.
The spacetime vev $b_\mu$ breaks diffeomorphisms.
The converse is also true.
If a nonscalar tensor vev on the spacetime manifold breaks diffeomorphisms,
then the associated local tensor will have a vev that breaks local
Lorentz symmetry.
In the case of a scalar,
the derivatives of the field will have vevs that
break local Lorentz symmetry.

Next, the question of how many NG modes there are and where they reside
was examined.  
Since there are six Lorentz symmetries and four diffeomorphisms,
which can all be broken when a tensor with a sufficient number of
indices acquires a fixed vev,
this means that in general up to ten NG modes can appear.
A general argument shows that these ten modes can all
be absorbed as additional degrees of freedom in the vierbein.
A simple counting argument supports this as well.
The vierbein has 16 components.
With Lorentz symmetry,
six of these modes can be gauged away.
They are usually chosen as the antisymmetric components.
Similarly, diffeomorphisms can be used to remove four additional
degrees of freedom.
This leaves six vierbein modes in the general case.
Einstein's theory has four of these modes as auxiliary,
resulting in only two massless modes for the graviton.
However,
in a more general gravitational theory,
there can be up to six propagating modes,
which in a vierbein formalism are the six vierbein degrees of freedom.
If Lorentz symmetry and diffeomorphisms are broken,
then the ability to gauge away some of the vierbein degrees of freedom is lost.
In particular,
since up to ten symmetries can be broken,
up to ten additional modes (the NG modes)
can appear in the vierbein.

The number of NG modes is also affected by the nature of the vev
and by the fact that the symmetry is a spacetime symmetry.
For example,
in the case of a vector vev,
which breaks three Lorentz symmetries and one diffeomorphism,
it might be expected that there would be three massless NG Lorentz modes
and one massless NG diffeomorphism mode.
However, in the case where the vector vev is a constant,
the diffeomorphism mode is found to be an auxiliary mode.
It is also found that there are only two propagating massless Lorentz modes.
The third Lorentz mode is found to be auxiliary as well.
In this case,
since the NG modes carry vector indices,
it makes sense that a massless vector would only have two propagating modes.
This clearly provides an example where the usual counting of NG modes
(one massless mode per broken generator) does not hold
for the case of a broken spacetime symmetry
\cite{counting}.

It was also found that the fate of the NG modes depends on the geometry.
In Riemann or Minkowski spacetime,
where the torsion is zero,
the NG modes appear as additional massless or auxiliary modes in the vierbein.
However,
in Riemann-Cartan spacetime,
which has nonzero torsion and where the spin connection has
degrees of freedom that are independent from the vierbein,
the possibility of a Higgs mechanism occurs.
This is because a mass term for the spin connection can form
when local Lorentz symmetry is spontaneously broken.
If the theory permits massless propagating modes for
the spin connection,
then these modes can acquire a mass.
In principle,
the mechanism is straightforward.
However,
finding a ghost-free unbroken model with a propagating
spin connection that is compatible with the mass term
is challenging.

A specific vector model with spontaneous Lorentz breaking,
called a bumblebee model,
has been used to illustrate the behavior of the NG modes.
For simplicity,
this overview will concentrate entirely on this example
for the case of a constant vev.
All of the general features described above will be applicable.

Bumblebee models in a gravitational theory were first looked
at by Kostelecky and Samuel as a simple model for investigating
the consequences of spontaneous Lorentz violation
\cite{ks2}.
Their properties have been studied in a variety of contexts
\cite{bbmodels}.
Much of the attention has focused on models with a
timelike vev.
It has been suggested that if a NG diffeomorphism mode
propagates in this case,
then it would have an unusual dispersion relation
\cite{ahclt}.

One especially noteworthy feature of the bumblebee model
occurs in Minkowski and Riemann spacetime.
It is found (in the linearized theory)
that the massless NG Lorentz modes behave essentially
as the photon in an axial gauge 
\cite{rbak05}.
Connections between Lorentz breaking and gauge fixing have been
noted previously,
leading to the suggestion that the photon is comprised of
NG modes due to spontaneous Lorentz breaking
\cite{dhfb,yn}.
However,
the approach of the bumblebee model is different.
It is not a $U(1)$ gauge theory,
since it contains a potential $V$ that is not $U(1)$ invariant.
The Lorentz breaking is therefore not a $U(1)$ gauge fixing choice.
Nonetheless,
the NG modes appear to behave at lowest order as photons in an axial gauge.
Moreover,
there are additional tell-tale signs of Lorentz breaking
\cite{rbak05}.
These include additional SME couplings in Riemann and Minkowski spacetime
as well as anomalous gravitational couplings in the case of a
Riemann geometry.
This offers the possibility of letting experiments determine whether
massless photons are the result of unbroken gauge symmetry or
whether they might be due to spontaneously broken Lorentz symmetry.

\subsection{Bumblebee Models}
\label{bbs}

The definition of a bumblebee model is that it is a vector theory 
in which the vector field $B^\mu$ acquires a nonzero vev,
which spontaneously breaks Lorentz symmetry.
The lagrangian consists of a kinetic term for $B^\mu$ and a
potential $V$ that induces spontaneous Lorentz breaking.
The potential is not $U(1)$ gauge invariant.
Typically,
the potential imposes a vev $b_a \ne 0$ for the vector in a local frame.
The vierbein relates this back to the spacetime vector as
$B_\mu = \vb \mu a b_a$.
For simplicity,
we assume a perturbative solution about a Minkowski background.
This permits us to drop the distinction between latin and Greek indices
and to write
\beq
\lvb \mu \nu = \et_{\mu\nu} + (\half h_{\mu\nu} + \ch_{\mu\nu}) ,
\label{vhchi}
\eeq
where the ten symmetric excitations $h_{\mu\nu} = h_{\nu\mu}$ are 
associated with the metric
$g_{\mu\nu} = \et_{\mu\nu} + h_{\mu\nu}$,
while the six antisymmetric components $\ch_{\mu\nu} = -\ch_{\nu\mu}$
are the local Lorentz degrees of freedom. 
In this background,
the vacuum solution takes the form
\beq
\vev{B^\mu} = b^\mu , \qquad \vev{e_{\mu\nu}} = \et_{\mu\nu} 
\quad .
\label{vac}
\eeq

There are a number of choices for the kinetic and potential terms.
Vector-current interactions and additional vector-curvature couplings 
that are forbidden in $U(1)$ gauge theory can be included as well
\cite{ks,akgrav}.

Here,
as an illustrative example,
we examine the model given by lagrangian
\bea
\cL_B &=& \fr 1 {2 \ka} (e R + \xi e B^\mu B^\nu R_{\mu\nu})
-\frac 1 4 e B_{\mu\nu} B^{\mu\nu} 
\nonumber \\
&& \quad\quad\quad
- e \la (B_\mu B^\mu \pm b^2) - e B_\mu J^\mu ,
\label{BBL}
\eea
where $\ka = 8 \pi G$ and $\xi$ is a coupling coefficient between
the vector field and the curvature.
The kinetic terms in this example are analogous to those
in Einstein-Maxwell theory.
However,
in the general case in a Riemann-Cartan spacetime,
the torsion contributes to these terms and the field strength is
defined by
\beq
B_{\mu\nu} = D_\mu B_\nu - D_\nu B_\mu ,
\label{Bmunu}
\eeq
where $D_\mu$ are covariant derivatives.
The potential term is 
\beq
V (B_\mu B^\mu \pm b^2) = \la (B_\mu B^\mu \pm b^2) ,
\label{Vsm}
\eeq
where $\la$ is a Lagrange-multiplier field.
It imposes the constraint that the vector field has a vev
$b^a$ obeying $b_a b^a = \mp b^2$
(with the sign corresponding to whether the vector is
timelike or spacelike).
The vector field can then be written in terms of
the vierbein and can be expanded perturbatively to give
\beq
B^\mu = \ivb \mu a b^a 
\approx b^\mu + (-\half h^{\mu\nu} + \ch^{\mu\nu}) b_\nu.
\label{Bv}
\eeq
The vierbein degrees of freedom include the NG modes.

This model can be studied in a linearized approximation.
The symmetric and antisymmetric components of the vierbein
transform as
\bea
h_{\mu\nu} 
&\rightarrow& 
h_{\mu\nu} ,
\nonumber\\
\ch_{\mu\nu} &\rightarrow& \ch_{\mu\nu} - \ep_{\mu\nu} ,
\label{LTs}
\eea
under infinitesimal Lorentz transformations,
while under infinitesimal diffeomorphisms
\bea
h_{\mu\nu} 
&\rightarrow& 
h_{\mu\nu} - \prt_\mu \xi_\nu - \prt_\nu \xi_\mu ,
\nonumber\\
\ch_{\mu\nu} 
&\rightarrow& 
\ch_{\mu\nu} - \half (\prt_\mu \xi_\nu - \prt_\nu \xi_\mu ) .
\label{chidiffeo}
\eea
In these expressions,
quantities of order 
$(\ep h)$, $(\ep \ch)$, $(\xi h)$, $(\xi \ch)$, 
etc.\ are assumed small 
and hence negligible in the linearized treatment.

The NG modes can be found as the virtual fluctuations
about the vacuum solution.
These can be written as
\beq
\de B^\mu = (B^\mu - b^\mu)
\approx (-\half h^{\mu\nu} + \ch^{\mu\nu}) b_\nu .
\label{name}
\eeq

It is useful to introduce projections on the transverse and
longitudinal components of $\de B^\mu$ along $b^\mu$.
Assuming $b^2 \ne 0$, these are given by
\beq
(P_{\parallel})^\mu_{\pt{\mu}\nu} = 
\fr {b^\mu b_\nu} {b^\si b_\si} \, , \quad
(P_{\perp})^\mu_{\pt{\mu}\nu}= 
\de^\mu_{\pt{\mu}\nu} 
- (P_{\parallel})^\mu_{\pt{\mu}\nu} .
\label{projs}
\eeq
Defining the projected fluctuations as
\beq
\cE^\mu = (P_{\perp})^\mu_{\pt{\mu}\nu} \de B^\nu
\, , \quad\quad
\rh^\mu = (P_{\parallel})^\mu_{\pt{\mu}\nu} \de B^\nu
\approx b^\mu \rh ,
\label{Xiproj}
\eeq
where 
\beq
\rh = - \fr {b^\mu h_{\mu\nu} b^\nu} {2 b^\si b_\si} .
\label{rho}
\eeq
lets us write the field $B^\mu$ as 
\beq
B^\mu \approx (1+\rh) b^\mu + \cE^\mu .
\label{Bprojs}
\eeq

In terms of these projections,
the NG Lorentz and diffeomorphism modes can be identified.
Under a virtual local particle Lorentz transformation
only components $\cE^\mu$ obeying $b_\mu \cE^\mu = 0$
are excited.
These are the NG Lorentz modes,
which evidently obey a condition similar to an axial-gauge condition
in $U(1)$ gauge theory.
If instead a virtual infinitesimal diffeomorphism is performed,
only the longitudinal component $\rh$ is excited.
It can therefore be identified as the NG diffeomorphism mode.
Note that a metric fluctuation about the vacuum solution,
\beq
\et_{\mu\nu} \rightarrow g_{\mu\nu}
\approx \et_{\mu\nu} - \prt_\mu \xi_\nu - \prt_\nu \xi_\mu ,
\label{gdiff}
\eeq
is generated by the diffeomorphism as well.

The dynamics of the NG modes depend on the background geometry.
Three cases corresponding to Minkowski, Riemann,
and Riemann-Cartan spacetime are examined in the following sections.

\subsection{Minkowski Spacetime}
\label{Mink}

In Minkowski spacetime,
the curvature and torsion equal zero, 
and the metric can be written as
\beq
g_{\mu\nu} = \et_{\mu\nu}.
\label{etg}
\eeq
The bumblebee lagrangian in Eq.\ \rf{BBL} reduces to
\beq
\cL_B = 
-\frac 1 4 B_{\mu\nu} B^{\mu\nu} 
- \la (B_\mu B^\mu \pm b^2) - B_\mu J^\mu .
\label{flatBB}
\eeq
In this case,
it is found that the diffeomorpism mode $\rh$
cancels in $B_{\mu\nu}$.
It is therefore an auxiliary mode and does not propagate.
The Lorentz modes are contained in the projection $\cE_\mu$.
Renaming this as $\cE_\mu \equiv A_\mu$ and calling
the field strength
$F_{\mu\nu} \equiv \prt_\mu A_\nu - \prt_\nu A_\mu$
lets us rewrite the lagrangian as
\beq
\cL_B \to \cL_{\rm NG} \approx 
-\frac 1 4 F_{\mu\nu} F^{\mu\nu} - A_\mu J^\mu 
- b_\mu J^\mu + b^\mu \prt_\nu \Xi_\mu J^\nu ,
\label{linflatBB}
\eeq
where $\Xi_\mu$ is the longitudinal diffeomorphism mode $\xi_\mu$
promoted to an NG field.
It is defined by $\rh = \partial_\mu \Xi^\mu$.
Note that varying with respoct to this auxiliary mode
yields the current-conservation law,
$\prt_\mu J^\mu = 0$.

The lagrangian $\cL_{\rm NG}$ is the effective quadratic lagrangian 
that governs the propagation of the NG modes in Minkowski space.
The field $A^\mu$ has three degrees of freedom and automatically
obeys an axial-gauge condition $b_\mu A^\mu = 0$.
It contains the three Lorentz NG modes.
Depending on the vev $b_\mu$,
the special cases of temporal gauge 
($A^0 = 0$) 
and pure axial gauge
($A^3 = 0$)
are possible. 

It can be seen that in Minkowski spacetime the NG modes 
resemble those of a massless photon
in $U(1)$ gauge theory in an axial gauge.
Unlike the gauge theory case,
however,
where the masslessness of the photon is due to
unbroken gauge symmetry,
in this case the masslessness of the photon
is a consequence of spontaneously broken Lorentz symmetry.
An important question is whether this interpretation of the
photon has experimentally verifiable consequences.
Clearly,
there is one additional interaction that does not
hold for the usual photon in gauge theory.
This is the Lorentz-violating term $b_\mu J^\mu$,
where $J^\mu$ is the charge current. 
This term can be identified with the SME term with coefficient $a^e_\mu$ 
that occurs in the QED limit of the SME
\cite{ck}.
This type of SME coefficient if it is constant
is known to be unobservable in 
experiments restricted to the electron sector
\cite{akgrav,ck,kleh}.
However,
it can generate signals in the quark 
and neutrino sectors.
Thus,
in experiments with multiple particle sectors,
the idea that the photon results from spontaneous Lorentz
violation can potentially be tested in Minkowski space.

\subsection{Riemann Spacetime}
\label{Riemann}

In Riemann geometry in a vierbein formalism,
the spin connection $\nsc \mu a b$ appears in covariant derivatives.
However,
the metric requirement,
\beq
D_\la \vb \mu a = 0,
\eeq
and the fact that the torsion vanishes
permits the spin connection to be completely determined
in terms of the vierbein as
\bea
\nsc \mu a b &=&
\half \uvb \nu a ( \prt_\mu \vb \nu b - \prt_\nu \vb \mu b)
- \half \uvb \nu b ( \prt_\mu \vb \nu a - \prt_\nu \vb \mu a)
\nonumber\\
&&
- \half \uvb \al a \uvb \be b \vb \mu c
(\prt_\al \lvb \be c - \prt_\be \lvb \al c).
\label{scvierb}
\eea
The spin connection has no independent degrees of freedom in
Riemann spacetime,
and the NG modes are still contained in the vierbein.
In this case (with gravity),
up to six of the 16 components of the vierbein 
can represent dynamical degrees of freedom associated 
with the gravitational fields.

We again consider the bumblebee lagrangian and vacuum as given in 
Eqs.\ \rf{BBL} and \rf{vac},
respectively.
The projector-operator decomposition of $B^\mu$ reveals that
there are four potential NG modes
contained in $\cE^\mu$ and $\rh$,
and the axial-gauge condition $b_\mu \cE^\mu = 0$ still holds
in Riemann spacetime.
The field strength $B_{\mu\nu}$ 
can be rewritten as
\beq
B_{\mu\nu} = (\prt_\mu \vb \nu a - \prt_\nu \vb \mu a) b_a ,
\label{vB}
\eeq
which suggests that the propagation of the vierbein is modified
by the bumblebee kinetic term.

The effective lagrangian for the NG modes 
can be found by expanding the bumblebee lagrangian 
to quadratic order,
keeping couplings to matter currents and curvature.
The result in terms of the decomposed fields is
\bea
\cL_{\rm NG} 
&\approx &
\fr 1 {2 \ka}[
e R 
+ \xi e b^\mu b^\nu R_{\mu\nu}
+ \xi e A^\mu A^\nu R_{\mu\nu} 
\nonumber \\
&& 
+ \xi e \rh(\rh + 2) b^\mu b^\nu R_{\mu\nu} 
+ 2\xi e (\rh + 1) b^\mu A^\nu R_{\mu\nu} 
]
\nonumber \\
&&
-\frac 1 4 eF_{\mu\nu} F^{\mu\nu} - eA_\mu J^\mu 
- eb_\mu J^\mu + eb^\mu \prt_\nu \Xi_\mu J^\nu ,
\nonumber \\
\label{NGcurv}
\eea
where again the Lorentz modes are relabeled as
$A_\mu \equiv \cE_\mu$,
which obeys $b_\mu A^\mu = 0$, 
and the field strength is 
$F_{\mu\nu} \equiv \prt_\mu A_\nu - \prt_\nu A_\mu$.
The gravitational excitations $h_{\mu\nu}$
obey the condition $h_{\mu\nu} b^\mu = 0$.

The form of this effective lagrangian
reveals that only two of the four potential NG modes propagate.
These are the transverse Lorentz NG modes.
The longitudinal Lorentz and the diffeomorphism NG modes
are auxiliary.
In particular,
the curvature terms do not provide kinetic terms for $\rh$.
This is because,
metric fluctuations in the form of a diffeomorphism excitation 
produce only a vanishing contribution 
to the curvature tensor at linear order.

In Riemann spacetime,
the NG Lorentz modes again resemble the photon in an axial gauge.
The interaction with the charged current $J_\mu$ also has
the appropriate form.
However,
possible signals for testing the idea that the photon
is due to Lorentz violation can be found.
In particular,
there are unconventional couplings 
of the curvature with $A^\mu$, $\rh$, and $b^\mu$.
The curvature couplings
$e A^\mu A^\nu R_{\mu\nu}$,
are forbidden by gauge invariance
in conventional Einstein-Maxwell electrodynamics,
but they can appear here in a theory with Lorentz violation.
The term $\xi e b^\mu b^\nu R_{\mu\nu}/2\ka$
corresponds to an SME coefficient of the $s^{\mu\nu}$ type
in the gravity sector of the SME.
The remaining terms also represent Lorentz-violating couplings
that are included in the SME.
Any of these signals could serve to provide experimental evidence 
for the idea that the photon is an NG mode 
due to spontaneous Lorentz violation.

\subsection{Riemann-Cartan Spacetime}
\label{RC}

In a Riemann-Cartan spacetime,
the vierbein $\vb \mu a$ 
and the spin connection 
$\nsc \mu a b$
are independent degrees of freedom.
As a result,
the effects of spontaneous Lorentz breaking
are very different from the cases of Minkowski
and Riemann spacetime.
In particular,
it has been found that when the torsion is nonzero
it is possible for a Higgs mechanism to occur
\cite{rbak05}.
This will be illustrated below in the context of the bumblebee
model in Riemann-Cartan spacetime.

One immediate question concerning the possibility of a Higgs mechanism in
a gravitational theory is whether the graviton acquires a mass or not.
Indeed,
even a small mass for the graviton
can modify the predictions of general relativity
leading to disagreement with experiment
\cite{vdvz}.
However,
it was shown some time ago that a conventional Higgs mechanism cannot 
give rise to a mass for the graviton since the 
terms that are generated involve derivatives of the metric
\cite{ks2}.

A generic lagrangian for a theory 
with spontaneous Lorentz violation
in Riemann-Cartan spacetime can be written as
\beq
\cL = \cL_0 
+ 
\cL_{\rm SSB}.
\label{th}
\eeq
Here,
we assume $\cL_0$ contains only gravitational terms
formed from the curvature and torsion and describes the unbroken theory,
while $\cL_{\rm SSB}$ induces spontaneous Lorentz violation.
For a Higgs mechanism to occur involving the spin connection,
$\cL_0$ should describe massless propagating modes for the spin connection
prior to the spontaneous breaking of Lorentz symmetry.
The theory should also be free of ghosts.
It turns out that these conditions 
severely restrict the possibilities for model building.
The number of ghost-free theories with 
massive and massless propagating spin connection modes is limited
\cite{svn,kf}.
The number of propagating modes in these models 
depends on the presence of additional accidental symmetries.
The symmetry-breaking lagrangian
$\cL_{\rm SSB}$ 
typically breaks one or more of the accidental symmetries 
when the tensor field acquires a vev,
which complicates the analysis of potential models.

In the bumblebee model in Eq.\ \rf{BBL} 
the symmetry-breaking part of the lagrangian is
\beq
\cl_{\rm SSB} = - \quar e B_{\mu\nu} B^{\mu\nu}
- e \la (B_\mu B^\mu \pm b^2) .
\label{Lssb}
\eeq
In a Riemann-Cartan spacetime,
the field strength $B_{\mu\nu}$ is defined in Eq.\ \rf{Bmunu}.
In terms of the vierbein and spin connection,
it becomes 
\beq
B_{\mu\nu} =
(\vb \mu b  \lulsc \nu a b - \vb \nu b  \lulsc \mu a b ) b_a .
\label{FS}
\eeq
Note that this expression reduces back to Eq.\ \rf{vB}
in the limits of Riemann and Minkowski spacetimes,
where the spin connection 
is given by Eq.\ \rf{scvierb}.

When $B_{\mu\nu}$ is squared,
quadratic terms in $\lulsc \mu a b$ appear in the lagrangian,
which perturbatively have the form
\beq
- \quar e B_{\mu\nu} B^{\mu\nu} 
\approx 
- \quar (\om_{\mu\rh\nu} - \om_{\nu\rh\mu})
(\om^{\mu\si\nu} - \om^{\nu\si\mu}) b^\rh b_\si .
\label{om2}
\eeq
It is these quadratic terms 
that suggest that a Higgs mechanism can occur
involving the absorption of the NG modes 
by the spin connection.
It should be noted that this is only possible in Riemann-Cartan spacetime
with nonzero torsion,
since otherwise (as in Riemann spacetime) the spin connection 
has no independent degrees of freedom.

In Ref.\ \cite{rbak05},
a number of different models for the kinetic terms $\cL_0$
were considered.
As mentioned,
the difficulty in building a viable model with a Higgs mechanism
comes from finding a kinetic term describing propagating modes
that are compatible with Eq.\ \rf{om2} as a mass term.
If ghosts are permitted,
this is straightforward.
For example, with the choice
\beq
\cl_{0} =  
\frac 1 4 R_{\la \ka \mu \nu} R^{\la \ka \mu \nu} .
\label{R2L}
\eeq
all the fields $\lsc \la \mu \nu$ with $\la \ne 0$
propagate as massless modes.
When this is combined with $\cl_{\rm SSB}$,
we find that among the propagating modes 
in the linearized theory there is a massive mode.
Other examples can be studied as well and are
aided by decomposing the fields $\lsc \la \mu \nu$
according to their spin-parity projections $J^P$
in three-dimensional space. 
This reveals that the mass term consists of a
physical $1^+$ mode and a $1^-$ gauge mode.
Models can be found in which
$\cL_0$ includes a massless $1^+$ mode.
However,
typically the propagating massless modes involve combinations of
$J^P$ projections,
which makes finding compatibility with $\cl_{\rm SSB}$
all the more challenging.

In the end,
a number of issues remain open for future investigation.
Studies of the large variety of possible Lorentz-invariant lagrangians $\cL_0$
can lead to new models in which
the spin connection acquires a mass due to spontaneous Lorentz breaking.
Different choices for $\cl_{\rm SSB}$ can also be considered,
including ones in which the spontaneous Lorentz violation
involves one or more tensor fields.
This would certainly affect the dynamics of the NG modes as well.
From a broader theoretical point of view,
the incorporation of spontaneous Lorentz violation in theories with torsion opens
up a new arena in the search for ghost-free models with propagating massive modes.

Certainly, there are implications for phenomenology in the context of
Riemann-Cartan spacetime.
The relevant mass scale in the Higgs mechanism 
is set by $b^2$.
Even if this is on the order of the Planck mass,
the existence of fields associated with Lorentz violation
could have effects on cosmology, black holes,
and gravitational radiation.
Since all of the relevant terms in any of these models
are included in the SME in Riemann-Cartan spacetime,
a systematic approach would be to investigate possible new signals in that context.

\section{Phenomenology}
\label{phenom}

The minimal SME described in Section \rf{msme} has been used
extensively in recent years by experimentalists and theorists
to search for leading-order signals of Lorentz violation.
To date,
Planck-scale sensitivity has been attained 
to the dominant SME coefficients in a number of experiments
involving different particle sectors.
These include experiments with
photons \cite{rcs,photonexpt1,photonexpt2,tobar05,schiller05,kmphotons,photonth},
electrons \cite{dehmelt,mitt,penn,hyd,etheory,eexpt2,eexpt3},
protons and neutrons \cite{ccexpt,Hmaser,cane,kl99,ccspace,ppenning},
mesons \cite{hadronexpt,akmesons},
muons \cite{hughes01,muong01,bkl00},
neutrinos \cite{ck,cg99,nuexpt,km04,nuth},
and the Higgs \cite{higgs}.
It should be noted that despite the length of this list of experiments,
a substantial portion of the SME coefficient space
remains unexplored.

In the remaining sections,
an overview of some of the recent tests of Lorentz and CPT symmetry
in a Minkowski background will be given.
In particular,
since many of the sharpest test are performed in high-precision
atomic and particle experiments involving photons and charged particles,
much of the focus will be on the QED limit of the minimal SME.
However,
two other particle sectors are briefly described as well.
These involve testing Lorentz and CPT symmetry with mesons and neutrinos.

\subsection{Mesons}
\label{mesons}

Experiments with mesons have long provided some of the sharpest tests of CPT.
Since CPT and Lorentz symmetry are intertwined in field theory,
these experiments also provide additional tests of Lorentz symmetry.
Investigations in the context of the SME have found very high sensitivity to
the CPT-odd $a_\mu$ coefficients in the SME.

The time evolution of a meson $P^0$ and its antimeson $\overline{P^0}$
is governed by a 2$\times$2 effective hamiltonian $\La$ 
in a description based on the Schr\"odinger equation.
Here,
$P$ represents one of the neutral mesons $K$, $D$, $B_d$, $B_s$.
The hamiltonian can be written as
\cite{akmesons,ll}
\beq
\La = 
\half \De\la
\left( \begin{array}{lr}
U + \xi 
& 
\quad VW^{-1} 
\\ & \\
VW \quad 
& 
U - \xi 
\end{array}
\right),
\label{uvwx}
\eeq
where the parameters $UVW\xi$ are complex.
The factor $\De\la/2$ 
ensures these parameters are dimensionless.
Imposing conditions on the trace and determinant gives the relations
$U \equiv \la/\De\la$
and $V \equiv \sqrt{1 - \xi^2}$.
The independent complex parameters 
$W = w \exp (i\om)$ and
$\xi = {\rm Re}[\xi] + i {\rm Im}[\xi]$
have four real components.
One is physically unobservable.
The argument $\om$ changes under a phase redefinition 
of the $P^0$ wave function.
The three others are physical.
The two real numbers
Re$[\xi]$, Im$[\xi]$
determine the amount of CPT violation,
with CPT preserved if and only if both are zero.

The dominant CPT-violating contributions to 
the effective hamiltonian $\La$
can be calculated as expectation values of interaction terms
in the SME.
The result in terms of $\xi$ is 
\beq
\xi \sim \be^\mu \De a_\mu
\quad ,
\label{dem}
\eeq
where $\be^\mu = \ga (1, \vec \be )$ is the four-velocity
of the $P$ meson in the laboratory frame
and the coefficients $\De a_\mu$
are combinations of SME coefficients.

The 4-velocity (and 4-momentum) dependence in Eq.\ \rf{dem} 
shows explicitly that CPT violation cannot be described 
with a constant complex parameter in quantum field theory
\cite{akmesons}.
Nonetheless, most experiments have fit their data to a constant
value of $\xi$.
Experiments in the kaon system
\cite{k99},
for example,
have attained bounds of order $10^{-4}$
on the real and imaginary parts of $\xi$.
More recently, however,
analyses have been performed taking into account that
in an experiment $\De a_\mu$ varies with the magnitude and
direction of the momentum and with sidereal time as
the Earth rotates.
These experiments have attained sensitivities to $\De a_\mu$
on the order of $10^{-20}$ GeV in the kaon system
and $10^{-15}$ GeV in the D system
\cite{hadronexpt}.
Additional bounds for the $B_d$ and $B_s$ systems
can be obtained as well in future analyses.

\subsection{Neutrinos}
\label{neutrinos}

A general analysis in the context of the SME has searched for possible
signals of Lorentz violation in neutrino physics
\cite{km04}.
Among other things,
it looked at how free neutrinos with Dirac and Majorana couplings
oscillate in the presence of Lorentz violation.
Remarkably,
a number of possible models exist in which Lorentz violation
(with or without massive neutrinos) contributes to neutrino oscillations.
One two-parameter model in particular,
consisting of massless neutrinos,
called the bicycle model,
reproduces features in observed data 
(except for the LSND experiment).
Indeed,
a statistical analysis performed using data from Super-Kamiokande
on atmospheric neutrinos finds that the fit based on the bicycle model
is essentially as good (within a small marginal error)
to the fit based on small mass differences
\cite{nuexpt}.
Further investigations looking for sidereal time variations
will be able to distinguish oscillations associated with Lorentz violation
from those due to small mass differences.

\subsection{QED Sector}
\label{qed}

Traditionally,
many of the sharpest tests of Lorentz and CPT symmetry have been made
with photons or in particle or atomic systems where the predominant 
interactions are described by QED.
This would include the original Michelson-Morley experiments and
their modern-day versions
\cite{photonexpt2,tobar05,schiller05}.
The Lorentz tests known as
Hughes-Drever experiments are atomic experiments in which two
high-precision atomic clocks consisting of different atomic species
are compared as the Earth rotates
\cite{ccexpt}.
These provide exceptionally sharp tests of Lorentz symmetry.
Similarly,
some of the best CPT tests for leptons and baryons 
-- involving direct comparisons of particles and antiparticles --
are made by atomic physicists working with Penning traps
\cite{dehmelt,mitt,ppenning}.

In order to look for the leading order signals of Lorentz and CPT
violation in these types of experiments,
it is useful to work with a
subset of the minimal SME lagrangian 
that is relevant to experiments in QED systems.
The QED limit of the minimal SME can be written as
\beq
{\mathcal L}_{\rm QED} = {\mathcal L}_0 + {\mathcal L}_{\rm int} 
\quad .
\label{lag}
\eeq
The lagrangian ${\mathcal L}_0$ contains the usual Lorentz-invariant
terms in QED describing photons, massive charged fermions, 
and their conventional couplings,
while ${\mathcal L}_{\rm int}$ contains the Lorentz-violating interactions.
Since the minimal SME in flat spacetime
is restricted to the remormalizable
and gauge-invariant terms in the full SME,
the QED sector interactions in ${\mathcal L}_{\rm int}$
have a finite number of terms.
For the case of photons and a single fermion species $\ps$
the Lorentz-violating terms are given by
\cite{note1}
\bea
{\mathcal L}_{\rm int} &=& - a_\mu \bar \ps \ga^\mu \ps
- b_\mu \bar \ps \ga _5 \ga^\mu \ps
+ ic_{\mu \nu} \bar \ps \ga^\mu D^\nu \ps 
\nonumber \\
&& \quad\quad
+ id_{\mu \nu} \bar \ps \ga_5 \ga^\mu D^\nu \ps 
- \half H_{\mu \nu} \bar \ps \si^{\mu \nu} \ps 
\nonumber \\
&&
-\frac14 (k_F)_{\ka\la\mu\nu} F^{\ka\la}F^{\mu\nu} 
+ \frac 12 (k_{AF})^\ka \ep_{\ka\la\mu\nu} A^{\la} F^{\mu\nu} 
\quad .
\label{qedsme}
\eea
Here,
$i D_\mu \equiv i \partial_\mu - q A_\mu$.
The terms with coefficients $a_\mu$, $b_\mu$ and $(k_{AF})_\mu$
are odd under CPT,
while those with
$H_{\mu \nu}$, $c_{\mu \nu}$, $d_{\mu \nu}$, and $(k_F)_{\ka\la\mu\nu}$
preserve CPT.
All seven terms break Lorentz symmetry.
In general,
superscript labels will be added to these parameters to
denote the particle species.

This lagrangian emerges naturally from the minimal SME sector for charged leptons,
following the usual assumptions of electroweak symmetry breaking and mass generation. 
Lagrangian terms of the same form are expected to describe protons and neutrons
in QED systems as well,
but where the SME coefficients represent composites
stemming from quark and gluon interactions.
It is certainly the case that QED and its relativistic quantum-mechanical
limits describe proton and neutron electromagnetic interactions in atoms in 
excellent agreement with experiments.
Defining terms involving composite SME parameters for protons and neutrons is 
therefore a reasonable extension of the theory.
The QED extension of the SME treats protons and neutrons
as the basic constituents of the theory.
The lagrangian ${\mathcal L}_{\rm int}$ then contains the most
general set of Lorentz-violating interactions in this context. 

Since the corrections due to Lorentz violation at low energy 
are known to be small,
it is sufficient in many situations to work in the context of relativistic
quantum mechanics using perturbation theory.
To do so,
a Hamiltonian is needed such that
\beq
i \partial_0 \ch = \hat H \ch 
\quad ,
\label{Heq}
\eeq
where $\hat H = \hat H_0 + \hat H_{\rm pert}$.
The perturbative hamiltonian $\hat H_{\rm pert}$
associated with Lorentz violation
can be generated using a Foldy-Wouthuysen approach and
by making appropriate field redefinitions
\cite{penn,kl99}.
The result for a massive fermion particle is
\bea
\hat H_{\rm pert} &=& a_\mu \ga^0 \ga^\mu 
- b_\mu \ga_5 \ga^0 \ga^\mu - c_{0 0} m \ga^0
- i (c_{0 j} + c_{j 0})D^j 
\nonumber \\
&&
+ i (c_{0 0} D_j - c_{j k} D^k) \ga^0 \ga^j
- d_{j 0} m \ga_5 \ga^j + i (d_{0 j} + d_{j 0}) D^j \ga_5
\nonumber \\
&&
+ i (d_{0 0} D_j - d_{j k} D^k) \ga^0 \ga_5 \ga^j
+ \half H_{\mu \nu} \ga^0 \si^{\mu \nu}
\quad .
\label{bigH}
\eea
Here, the letters $j,k,l$, etc.\ represent the three spatial directions
in a laboratory frame.
The $j=3$ (or $z$ direction) is usually chosen as the quantization axis.
The corresponding hamiltonian for the antiparticle can
be obtained using charge conjugation.

The SME coefficients are expected to be fixed with respect to a nonrotating
coordinate frame.
As a result,
the SME coefficients $b_0$, $b_j$, etc.\ would change as the Earth moves.
In order to give measured bounds in a consistent manner,
a nonrotating frame is chosen.
Often,
this is chosen as a sun-centered frame using celestial equatorial coordinates.
These are denoted using upper-case letters $T$,$X$,$Y$,$Z$.
Typically,
experiments sensitive to sidereal time variations are sensitive to
a combination of parameters,
which are denoted using tildes.
For example,
the $b_\mu$ tilde coefficients with $\mu=j$ are defined as
\beq
\tilde b_j^e  \equiv b_j^e - m d_{j0}^e - \half \ve_{jkl} H_{kl}^e 
\quad ,
\label{btilde}
\eeq
These combinations are projected onto the nonrotating frame,
where the components 
with respect to the celestial equatorial coordinate frame are
$b^e_X$, $b^e_Y$, $b^e_Z$, etc.
The relation between the laboratory and nonrotating components is
\bea
\tilde b_1^e 
&=& 
\tilde b^e_X \cos \ch \cos \Om t
+ \tilde b^e_Y \cos \ch \sin \Om t - \tilde b^e_Z \sin \ch ,
\nonumber\\
\tilde b_2^e 
&=& 
- \tilde b^e_X \sin \Om t + \tilde b^e_Y \cos \Om t ,
\nonumber\\
\tilde b_3^e 
&=& 
\tilde b^e_X \sin \ch \cos \Om t
+ \tilde b^e_Y \sin \ch \sin \Om t + \tilde b^e_Z \cos \ch .
\quad 
\label{map}
\eea
The angle $\ch$ is between the $j=3$ lab axis and the direction
of the Earth's rotation axis,
which points along $Z$.
The angular frequency $\Om \simeq {2 \pi}/{({\rm 23 h \, 56 m})}$
is that corresponding to a sidereal day.

\section{Tests in QED}
\label{qedtests}

Before examining individual tests of Lorentz symmetry in QED systems,
it is useful to examine some of the more general
results that have emerged from these investigations.
One general feature is that
sensitivity to Lorentz and CPT violation in these experiments
stems primarily from their
ability to detect very small anomalous energy shifts.
While many of the experiments were originally designed to
measure specific quantities,
such as charge-to-mass ratios of particles and antiparticles
or differences in $g$ factors,
it is now recognized that these experiments are most effective as
Lorentz and CPT tests when all of the energy levels in the system
are investigated for possible anomalous shifts.
As a result of this,
a number of new signatures of Lorentz and CPT violation have been
discovered in recent years that were overlooked previously.

A second general feature concerns how these atomic experiments are 
typically divided into two groups.
The first (Lorentz tests) looks for sidereal time
variations in the energy levels of a particle or atom.
The second (CPT tests) looks for a difference in
the energy levels between a particle (or atom) and its
antiparticle (or antiatom).
What has been found is that the
sensitivity to Lorentz and CPT violation in these
two classes of experiments is not distinct.
Experiments traditionally viewed as Lorentz tests are
also sensitive to CPT symmetry and vice versa.
Nonetheless,
it is important to keep in mind that 
that there are differences as well.
For example,
the CPT experiments comparing
matter and antimatter are directly sensitive to CPT-violating
parameters, such as $b_\mu$,
whereas Lorentz tests are sensitive to combinations of 
CPT-preserving and CPT-violating parameters,
which are denoted using a tilde.
Ultimately,
both clases of experiments are important and
should be viewed as complementary.

It has become common practice to express
sensitivities to Lorentz and CPT violation 
in terms of the SME coefficients.
This provides a straightforward approach that allows comparisons across different
types of experiments.
Since each different particle sector in the QED extension
has an independent set of Lorentz-violating SME coefficients,
these are distinguished using superscript labels.
A thorough investigation of Lorentz and CPT violation necessarily
requires looking at as many different particle sectors as possible.

\subsection{Photons}
\label{photons}

The lagrangian describing a freely propagating
photon in the presence of Lorentz violation is given by
\cite{kmphotons}
\beq
\cl = -\frac14 F_{\mu\nu}F^{\mu\nu}
-\frac14 (k_F)_{\ka\la\mu\nu}
     F^{\ka\la}F^{\mu\nu}\ 
+ \frac12 (k_{AF})^\ka \ep_{\ka\la\mu\nu} A^{\la} F^{\mu\nu} ,
\eeq
where the field strength $F_\mn$ is defined by
$F_\mn \equiv \prt_\mu A_\nu -\prt_\nu A_\mu$.

The coefficient $k_{AF}$,
which is odd under CPT, 
has been investigated extensively both theoretically and experimentally
\cite{photonexpt1,kmphotons}.
Theoretically, it is found that this term leads to negative-energy contributions 
and is a potential source of instability in the theory.  
One solution is to set $k_{AF}$ to zero,
which has been shown to be consistent with radiative corrections in the SME.
However,
stringent experimental constraints also exist
consistent with $k_{AF} \approx 0$. 
These result from studying the polarization of radiation
from distant radio galaxies.
In what follows,
we will therefore ignore the effects of the $k_{AF}$ terms.

The terms with coefficients $k_{F}$,
which is even under CPT, have been
investigated more recently
\cite{kmphotons}.
These terms provide positive-energy contributions.
There are 19 independent components in the
$k_{F}$ coefficients.
It is useful to rewrite them in terms of a new set,
$\tilde\ka_{e+}$, $\tilde\ka_{e-}$,
$\tilde\ka_{o+}$, $\tilde\ka_{o-}$, and $\tilde\ka_{\rm tr}$.  
Here, $\tilde\ka_{e+}$, $\tilde\ka_{e-}$, and $\tilde\ka_{o-}$
are $3\times3$ traceless symmetric matrices 
(with 5 independent components each),
while $\tilde\ka_{o+}$ is a
$3\times3$ antisymmetric matrix 
(with 3 independent components),
and the remaining coefficient $\tilde\ka_{\rm tr}$
is the only rotationally invariant component.

The lagrangian can be written in terms of the
new set and the usual electric and magnetic fields
$\vec E$ and $\vec B$  as follows:
\bea
\cl &=& \half [(1+\tilde\ka_{\rm tr})\vec E^2
-(1-\tilde\ka_{\rm tr})\vec B^2]
+\half \vec E\cdot(\tilde\ka_{e+}
+\tilde\ka_{e-})\cdot\vec E
\nonumber \\
&&
-\half\vec B\cdot(\tilde\ka_{e+}
-\tilde\ka_{e-})\cdot\vec B 
+\vec E\cdot(\tilde\ka_{o+}
+\tilde\ka_{o-})\cdot\vec B\ 
\quad .
\label{modL}
\eea

This lagrangian gives rise to modifications of Maxwell's equations,
which have been explored in recent astrophysical 
and laboratory experiments.
Ten of the coefficients,
$\tilde\ka_{e+}$ and $\tilde\ka_{o-}$, 
lead to birefrigence of light.
Bounds on these parameters of order $2 \times 10^{-32}$ have been obtained
from spectropolarimetry of light from distant galaxies 
\cite{kmphotons}.
The nine coefficients,
$\tilde\ka_{\rm tr}$, $\tilde\ka_{e-}$, and $\tilde\ka_{o+}$,
have been bounded in a series of recent laboratory  photon experiments.
Seven of the eight $\tilde\ka_{e-}$ and $\tilde\ka_{o+}$ coefficients,
have been bounded in experiments using optical and microwave cavities. 
Sensitivities on the order of $\tilde\ka_{o+} \lsim 10^{-11}$ 
and $\tilde\ka_{e-} \lsim 10^{-15}$ have been attained
\cite{photonexpt2}.
The trace coefficient has been estimated to have an upper
bound of $\tilde\ka_{\rm tr} \lsim 10^{-4}$ from Ives-Stilwell experiments
\cite{tobar05}.
The remaining $\tilde\ka_{e-}$ coefficient has recently been
bounded at the level of $10^{-14}$ using a rotating apparatus
\cite{schiller05}.

\subsection{Penning Traps}
\label{penning}

There are primarily two leading-order signals of Lorentz and CPT violation 
that can be searched for in experiments in Penning traps 
\cite{penn}.
One is a traditional CPT test,
comparing particles and antiparticles,
while the other is a Lorentz test that looks for sidereal time variations.
Both types of signals have been investigated in recent years in experiments with
electrons and positrons.
The experiments involve making high-precision measurements of the
anomaly frequency $\om_a$ and the cyclotron frequency $\om_c$
of the trapped electrons and/or positrons.

The first test was a reanalysis was performed by Dehmelt's group using existing
data for electrons and positrons in a Penning trap
\cite{dehmelt}.
The idea was to look for an instantaneous difference in the
anomaly frequencies of electrons and positrons,
which can be nonzero when CPT and Lorentz symmetry are broken.
Dehmelt's original measurements of $g-2$
did not involve looking for possible instantaneous variations in $\om_a$.
Instead,
the ratio $\om_a/\om_c$ was computed using averaged values.
However, 
Lorentz-violating corrections to the 
anomaly frequency $\om_a$ can occur even if the $g$ factor remains unchanged.
An alternative analysis therefore looks for an
instantaneous difference in the electron and positron anomaly frequencies.
The new bound found by Dehmelt's group can be expressed in terms of the parameter $b^e_3$,
which is the component of $b^e_\mu$ along the quantization
axis in the laboratory frame.
The bound they obtained is $|b^e_3| \lsim 3 \times 10^{-25}$ GeV.

The second signal for Lorentz and CPT violation in the electron
sector involves measurements of the electron alone
\cite{mitt}.
Here,
the idea is that the Lorentz and CPT-violating interactions depend on
the orientation of the quantization axis in the laboratory frame,
which changes as the Earth turns on its axis.
As a result,
both the cyclotron and anomaly frequencies have small corrections which
cause them to exhibit sidereal time variations.
Such a signal can be measured using just electrons,
which eliminates the need for comparison with positrons.
The bounds in this case are given with respect to a
nonrotating coordinate system such as celestial equatorial coordinates.
The interactions involve a combination of laboratory-frame components
that couple to the electron spin.
The combination is denoted as
$\tilde b_3^e  \equiv b_3^e - m d_{30}^e - H_{12}^e$.
The bound can be expressed in terms of components $X$, $Y$, $Z$ 
in the nonrotating frame.
It is given as
$|\tilde b_J^e| \lsim 5 \times 10^{-25} {\rm GeV}$ for $J=X,Y$.

Although no $g-2$ experiments have been made for protons or antiprotons,
there have been recent bounds obtained on Lorentz violation in comparisons
of cyclotron frequencies of antiprotons and $H^-$ ions confined in a Penning trap
\cite{ppenning}.
In this case the sensitivity is to the dimensionless parameters $c^p_{\mu \nu}$.
Future experiments with protons and antiprotons will be able to
provide tests that are sensitive to $b^p_{\mu}$

\subsection{Clock-Comparison Experiments}
\label{ccexp}

The classic Hughes-Drever experiments
are atomic clock-comparison tests of Lorentz invariance
\cite{ccexpt,kl99}.
There have been a number of different types of these
experiments performed over the years,
with steady improvements in their sensitivity.
They involve making high-precision comparisons of 
atomic clock signals as the Earth rotates.
The clock frequencies are typically hyperfine or Zeeman transitions.
Many of the sharpest Lorentz bounds for the proton, neutron, and electron
stem from atomic clock-comparison experiments.
For example,
Bear {\it et al.} in 
Ref.\ \cite{ccexpt}
used a two-species noble-gas maser to
test for Lorentz and CPT violation in the neutron sector.
They obtain a bound
$|\tilde b_J^n| \lsim 10^{-31} {\rm GeV}$ for $J=X,Y$,
which is currently the best bound for the neutron sector.

It should also be pointed out that certain assumptions about the nuclear
configurations must be made to obtain bounds in
clock-comparison experiments.
For this reason,
these bounds should be viewed as good to within about
an order of magnitude.
To obtain cleaner bounds it is necessary to consider
simpler atoms or to perform more sophisticated nuclear modeling.

Note as well that these Earth-based laboratory experiments are not 
sensitive to Lorentz-violation coefficients along the $J=Z$
direction parallel to Earth's rotation axis.  
They also neglect the velocity effects due to Earth's 
motion around the sun,
which would lead to bounds on the timelike components along $J=T$.
These limitations can be overcome by performing experiments in space
or by using a rotation platform.
The earth's motion can also be taken into account.
A recent boosted-frame analysis of the dual noble-gas maser
experiment has yielded bounds on the order of $10^{-27}$ GeV
on many boost-dependent SME coefficients for the neutron
that were previously unbounded
\cite{cane}.

\subsection{Experiments in Space}
\label{space}

Clock-comparison experiments performed in space 
would have several advantages over traditional
ground-based experiments 
\cite{ccspace}.
For example,
a clock-comparison experiment conducted aboard 
the International Space Station (ISS)
would be in a laboratory frame that is both rotating and boosted.
It would therefore immediately gain sensitivity to
both the $Z$ and timelike directions.
This would more than triple the number of Lorentz-violation
parameters that are accessible in a clock-comparison experiment.
Another advantage of an experiment aboard the ISS is
that the time needed to acquire data would be greatly
reduced (by approximately a factor of 16).
In addition,
new types of signals would emerge that have no
analogue in traditional Earth-based experiments.
The combination of these advantages should result in
substantially improved limits on Lorentz and CPT violation.
Unfortunately,
the USA has canceled its missions aimed at testing fundamental
physics aboard the ISS.
However,
there is still a European mission planned for
the ISS which will compare atomic clocks and H masers.
Therefore,
the opportunity to perform these new Lorentz and CPT
tests is still a possibility.

\subsection{Hydrogen and Antihydrogen}
\label{hyd}

Hydrogen atoms have the simplest nuclear structure,
and antihydrogen is the simplest antiatom.
These atoms (or antiatoms) therefore provide opportunities
for conducting especially clean Lorentz and CPT tests
involving protons and electrons.

There are three experiments underway at CERN that 
can perform high-precision Lorentz and CPT tests in antihydrogen
\cite{cpt123}.
Two of the experiments (ATRAP and ATHENA) intend to
make high-precision spectroscopic measurements of the 1S-2S
transitions in hydrogen and antihydrogen.
These are forbidden (two-photon) transitions that have a relative linewidth
of approximately $10^{-15}$.
The ultimate goal is to measure the line center of this
transition to a part in $10^3$ yielding a frequency comparison
between hydrogen and antihydrogen at a level of $10^{-18}$.
An analysis of the 1S-2S transition in the context of the
SME shows that the magnetic field plays an important role
in the attainable sensitivity to Lorentz and CPT violation
\cite{hyd}.
For instance,
in free hydrogen in the absence of a magnetic field,
the 1S and 2S levels are shifted by equal amounts at leading order.
As a result,
in free H or $\bar {\rm H}$ there are no leading-order corrections 
to the 1S-2S transition frequency.
In a magnetic trap,
however,
there are fields that can mix the spin states in the
four different hyperfine levels.
Since the Lorentz-violating interactions depend on the spin orientation,
there will be leading-order sensitivity
to Lorentz and CPT violation in comparisons of 1S-2S transitions in
trapped hydrogen and antihydrogen.
At the same time,
however,
these transitions are field-dependent,
which creates additional experimental challenges 
that would need to be overcome.

An alternative to 1S-2S transitions is to consider the sensitivity
to Lorentz violation in ground-state Zeeman hyperfine transitions.
It is found that there are leading-order corrections in these levels
in both hydrogen and antihydrogen
\cite{hyd}.
The ASACUSA group at CERN is planning to measure the Zeeman hyperfine 
transitions in antihydrogen.
Such measurements will provide a direct CPT test.

Experiments with hydrogen alone have been performed using a maser
\cite{Hmaser}.
They attain exceptionally sharp sensitivity to Lorentz and CPT 
violation in the electron and proton sectors of the SME.
These experiments use a double-resonance technique that does
not depend on there being a field-independent point for the transition.
The sensitivity for the proton attained in these experiments 
is $|\tilde b_J^p| \lsim 10^{-27}$ GeV.
Due to the simplicity of hydrogen,
this is an extremely clean bound and is currently the most stringent test
of Lorentz and CPT violation for the proton.

\subsection{Muon Experiments}
\label{muons}

Experiments with muons involve second-generation leptons and
provide tests of CPT and Lorentz symmetry that are independent
of the tests involving electrons.
There are several different types of experiments with muons
that have recently been conducted,
including muonium experiments
\cite{hughes01}
and $g-2$ experiments with muons at Brookhaven
\cite{muong01}.
In muonium,
experiments measuring the frequencies
of ground-state Zeeman hyperfine transitions
in a strong magnetic field have the greatest sensitivity
to Lorentz and CPT violation.
A recent analysis has searched for sidereal time variations
in these transitions.
A bound at the level of $| \tilde b^\mu_J| \le 2 \times 10^{-23}$ GeV 
has been obtained
\cite{hughes01}.
In relativistic $g-2$ experiments using positive muons
with ``magic'' boost parameter $\de = 29.3$,
bounds on Lorentz-violation parameters are possible at
a level of $10^{-25}$ GeV.
However,
the analysis of these experiments is still underway at Brookhaven.

\subsection{Spin Polarized Torsion Pendulum}
\label{torpend}

Experiments using spin polarized torsion pendula have been conducted 
at the University of Washington and in Taiwan. 
These experiments currently provide the sharpest bounds
on Lorentz and CPT symmetry in the electron sector
\cite{eexpt2}.
These experiments are
able to achieve very high sensitivity to
Lorentz violation because the torsion pendula have a huge
number of aligned electron spins but a negligible magnetic field. 

The pendulum at the University of Washington
is built out of a stack of toroidal magnets,
which in one version of the experiment achieved
a net electron spin $S \simeq 8 \times 10^{22}$.
The apparatus is suspended on a rotating turntable and  
the time variations of the twisting pendulum are measured.
An analysis of this system shows that in addition to a signal having the
period of the rotating turntable,
the effects due to Lorentz and CPT violation also cause additional
time variations with a sidereal period caused by the rotation
of the Earth.
The group at the University of Washington has analyzed data taken in 1998
and find that thay have sensitivity to the electron coefficients
at the levels of $|\tilde b_J^e| \lsim 10^{-29}$ GeV for $J=X,Y$ and
$|\tilde b_Z^e| \lsim 10^{-28}$ GeV.
More recently,
a new pendulum has been built,
and it is expected that 20-fold improved sensitivities will
be attained
\cite{bh05}.

The Taiwan experiment also uses a rotating torsion pendulum,
which is made of a ferrimagnetic material.
This group achieved a net polarization of
$S \simeq 8.95 \times 10^{22}$ electrons in their pendulum.
The bounds they obtain  for the electron are
at the levels of $|\tilde b_J^e| \lsim 3.1 \times 10^{-29}$ GeV for $J=X,Y$ and
$|\tilde b_Z^e| \lsim 7.1 \times 10^{-28}$ GeV.

\section{Conclusions}
\label{concl}

This overview describes the development and use of the SME
as the theoretical framework describing Lorentz violation in the
context of field theory.
The philosophy of the SME is that any interactions that
are observer invariant and involve known fields at low energy
are included in the theory.
As an incremental first step,
the minimal SME (and its QED limit) 
can be constructed.
This theory maintains gauge invariance and power-counting
renormalizability.
It is the suitable framework for investigating leading-order
signals of Lorentz violation.

In addition to constructing the SME,
we have examined the special case of spontaneous Lorentz breaking.
In particular,
the question of what the fate of the Nambu-Goldstone modes is
when Lorentz symmetry is spontaneously broken has been addressed.
We have demonstrated that spontaneous particle Lorentz violation 
is accompanied by spontaneous particle diffeomorphism violation
and vice versa,
and that up to 10 NG modes can appear.
These modes can comprise 10 of the 16 modes of the vierbein
that in a Lorentz-invariant theory are 
gauge degrees of freedom.
The fate of the NG modes
is found to depend also on the spacetime geometry 
and on the behavior of the tensor vev inducing spontaneous 
Lorentz violation.
These results have been illustrated using a bumblebee model.
In Minkowski and Riemann spacetimes,
it is found that the NG modes propagate like the photon 
in an axial gauge.
In Riemann-Cartan spacetimes,
the interesting possibility exists 
that the spin connection could absorb the propagating NG modes 
in a gravitational version of the Higgs mechanism.
This unique feature of gravity theories with torsion
may offer another phenomenologically viable route 
for constructing realistic models
with spontaneous Lorentz violation. 

Phenomenology has been investigated using the minimal SME.
Experiments in QED systems continue to provide many of the
sharpest tests of Lorentz and CPT symmetry.
In recent years,
a number of new astrophysical and laboratory 
tests have been performed that have lead
to substantially improved sensitivities for the photon.
Similarly,
atomic experimentalists continue to find ways of improving 
the sensitivity to Lorentz violation in many of the matter sectors of the SME.
For comparison across different atomic experiments a
summary of recent bounds on the $\tilde b_J$ coefficients
in the minimal SME is given in Table 1.
These bounds are within the range of sensitivity associated
with suppression factors arising from the Planck scale.
A more complete table would list all of the
coefficients in the minimal SME.
Note that many SME coefficients have still not been measured.
Future experiments, 
in particular those performed in boosted frames, 
are likely to provide sensitivity to many of these
currently unmeasured SME coefficients.
In addition,
the overall sensitivity of these experiments is
expected to improve over the coming years.

\begin{table}[t]
\begin{center}
%\footnotesize
% =========================== start table ==================
\noindent
\renewcommand{\arraystretch}{1.2}
\begin{tabular}{|c|c|c|c|}
\hline\hline
Expt & Sector & Params ($J=X,Y)$ & Bound (GeV)
\\
\hline\hline
Penning Trap & electron & $\tilde b_J^e$ & $5 \times 10^{-25}$ \\[2mm]
\cline{1-4}
Hg-Cs clock  & electron & $\tilde b_J^e$ & $\sim 10^{-27}$ \\[2mm]
\cline{2-4}
comparison & proton & $\tilde b_J^p$ & $\sim 10^{-27}$ \\[2mm]
\cline{2-4}
 & neutron & $\tilde b_J^n$ & $\sim 10^{-30}$ \\[2mm]
\cline{1-4}
He-Xe dual maser & neutron & $\tilde b_J^n$ & $\sim 10^{-31}$ \\[2mm]
\cline{1-4}
H maser & electron & $\tilde b_J^e$ & $10^{-27}$ \\[2mm]
\cline{2-4}
 & proton & $\tilde b_J^p$ & $10^{-27}$ \\[2mm]
\cline{1-4}
Muonium & muon & $\tilde b_J^\mu$ & $2 \times 10^{-23}$ \\[2mm]
\cline{1-4}
Spin Pendulum & electron & $\tilde b_J^e$ & $10^{-29}$ \\[2mm]
&& $\tilde b_Z^e$ & $10^{-28}$ \\[2mm]
\hline
\hline
\end{tabular}
%\normalsize
% ============================ end table ===================
\renewcommand{\arraystretch}{1.0}
\caption{Summary of leading-order bounds for the parameter $\tilde b_J$.}
\vspace{0.2cm}
\end{center}
\end{table}

\vfill\eject

%\printindex

\begin{thebibliography}{99}
\bibitem{note1}
Additional terms that are forbidden by the requirements of gauge invariance
and renormalizability in the minimal SME can arise effectively in a QED
extension due to strong binding in a nucleus.
These coefficients are labeled as $e_\mu$, $f_\mu$, and $g_{\la\mu\nu}$.
The are included in the general investigations described in
Ref.\ \cite{kl99}.
However,
for simplicity,
they are ignored here.
\end{thebibliography}
\end{document}